\documentclass[sigconf, screen, authorversion]{acmart}
\AtBeginDocument{%
  \providecommand\BibTeX{{%
    \normalfont B\kern-0.5em{\scshape i\kern-0.25em b}\kern-0.8em\TeX}}}

\usepackage{array, booktabs, tabularx,  multirow}

\usepackage{caption}
\usepackage{subcaption}

\definecolor{navy}{RGB}{52, 85, 139}

\newcommand{\revise}[1]{{#1}}
\newcommand{\italic}[1]{{#1}}

\definecolor{venuec}{HTML}{ef8d22}
\definecolor{defaultc}{HTML}{e5e5e5}

\copyrightyear{2022} 
\acmYear{2022} 
\setcopyright{acmcopyright}\acmConference[CHI '22]{CHI Conference on Human Factors in Computing Systems}{April 29-May 5, 2022}{New Orleans, LA, USA}
\acmBooktitle{CHI Conference on Human Factors in Computing Systems (CHI '22), April 29-May 5, 2022, New Orleans, LA, USA}
\acmPrice{15.00}
\acmDOI{10.1145/3491102.3502048}
\acmISBN{978-1-4503-9157-3/22/04}

\begin{document}

\title{Structure-aware Visualization Retrieval}
\vspace{2em}

% \affiliation{%

\author{Haotian Li}
\affiliation{%
  \institution{The Hong Kong University of Science and Technology}
  \city{Hong Kong SAR}
  \country{China}
}
\affiliation{%
  \institution{Singapore Management University}
  \country{Singapore}
}
\email{haotian.li@connect.ust.hk}

\author{Yong Wang}
\affiliation{%
  \institution{Singapore Management University}
  \country{Singapore}
}
\email{yongwang@smu.edu.sg}

\author{Aoyu Wu}
\affiliation{%
  \institution{The Hong Kong University of Science and Technology}
  \city{Hong Kong SAR}
  \country{China}
}
\email{awuac@connect.ust.hk}

\author{Huan Wei}
\affiliation{%
  \institution{The Hong Kong University of Science and Technology}
  \city{Hong Kong SAR}
  \country{China}
}
\email{hweiad@connect.ust.hk}

\author{Huamin Qu}
\affiliation{%
  \institution{The Hong Kong University of Science and Technology}
  \city{Hong Kong SAR}
  \country{China}
}
\email{huamin@cse.ust.hk}

% \renewcommand{\shortauthors}{Li et al.}

\begin{abstract}
With the wide usage of data visualizations, a huge number of Scalable Vector Graphic~(SVG)-based visualizations have been created and shared online. Accordingly, there has been an increasing interest in exploring how to retrieve \revise{perceptually} similar visualizations from a large corpus, since it can benefit various downstream applications \revise{such as} visualization recommendation. Existing methods mainly focus on the visual \revise{appearance} of visualizations by regarding them as \revise{bitmap images}. However, the structural information intrinsically existing in SVG-based visualizations is ignored. Such structural information can delineate the spatial and hierarchical relationship among visual elements, and characterize visualizations thoroughly from a new perspective. This paper presents a structure-aware method to advance the performance of visualization retrieval by collectively considering both the visual and structural information. We extensively evaluated our approach through quantitative comparisons, a user study and case studies. The results demonstrate the effectiveness of our approach and its advantages over existing methods.
\end{abstract}

\begin{CCSXML}
<ccs2012>
   <concept>
       <concept_id>10003120.10003145</concept_id>
       <concept_desc>Human-centered computing~Visualization</concept_desc>
       <concept_significance>500</concept_significance>
       </concept>
   <concept>
       <concept_id>10002951.10003317</concept_id>
       <concept_desc>Information systems~Information retrieval</concept_desc>
       <concept_significance>300</concept_significance>
       </concept>
   <concept>
       <concept_id>10010147.10010257</concept_id>
       <concept_desc>Computing methodologies~Machine learning</concept_desc>
       <concept_significance>300</concept_significance>
       </concept>
 </ccs2012>
\end{CCSXML}

\ccsdesc[500]{Human-centered computing~Visualization}
\ccsdesc[300]{Information systems~Information retrieval}
\ccsdesc[300]{Computing methodologies~Machine learning}

\keywords{Data Visualization,  Visualization Retrieval, Visualization Similarity, Representation Learning, Visualization Embedding}


\maketitle

\section{Introduction}
Data visualization provides users with a powerful approach to analyze enormous data, communicate insights and achieve efficient decision-making.
Along with the popularity of visualizations, a huge number of visualizations based on Scalable Vector Graphics~(SVGs) have been created and shared online.
Compared with bitmap-based visualizations, SVG-based visualizations have many advantages such as the support of interactions~\cite{battle2018beagle} and quality-preserving resizing.
Thus, SVGs have been adopted by various online platforms to store and present visualizations, for example, Plotly\footnote{\url{https://plotly.com/}} and Observable\footnote{\url{https://observablehq.com/}}.
With such a large volume of visualizations online,
how to retrieve similar visualizations has attracted growing research interest from both academia and industry~\cite{opperman2021vizsnippets,oppermann2021vizcommender,saleh2015similarity} due to its significant importance for many downstream tasks. 
Specifically, the retrieval of similar visualizations is fundamental to downstream tasks such as \revise{creating visualization collections~\cite{opperman2021vizsnippets} and recommending visualizations~\cite{oppermann2021vizcommender}.}

To achieve effective retrieval of \revise{similar} visualizations, the core problem is to characterize the similarity between two visualizations.
\revise{Existing studies mainly focus on estimating the similarity between visualizations according to the data or perceptual similarity.
The existing methods based on \italic{\textit{data similarity}}~\cite{siddiqui2016effortless, vartak2015seedb, siddiqui2020shapesearch} focus on the characteristics of data such as data distribution,
while the visual appearance of visualizations is often ignored.
\italic{\textit{Perceptual similarity}} mainly refers to the similarity of visualizations perceived by users, which can also reflect the data similarity.
Since the original data is not always available with the visualizations, 
the application of visualization retrieval methods based on data similarity is quite limited.
Compared with the direct computation of data similarity, the computation of perceptual similarity does not rely on the original data. 
To compute the perceptual similarity, existing approaches~\cite{saleh2015similarity, ma2020scatternet, zhang2021chartnavigator} first extract the visual feature vectors from visualizations and further calculate the distance between feature vectors to measure their similarity.
\revise{These methods
mainly extract the visual features of visualizations at the level of pixels.}
For example, Saleh et al.~\cite{saleh2015similarity} measured the visualization similarity by using the color distribution of different pixels~(i.e., color histograms).
\revise{Recently, deep learning-based methods~\cite{ma2020scatternet, zhang2021chartnavigator} have been proposed to extract visual features automatically by treating visualizations as bitmap images (e.g., the images in ImageNet~\cite{deng2009imagenet})}. 
However, few prior studies have considered the structural information of visualizations that exists in SVGs by nature, 
when characterizing the perceptual similarity of visualizations.}

\begin{figure*}
     \centering
     \begin{subfigure}[b]{0.33\linewidth}
         \centering
         \includegraphics[width=\linewidth]{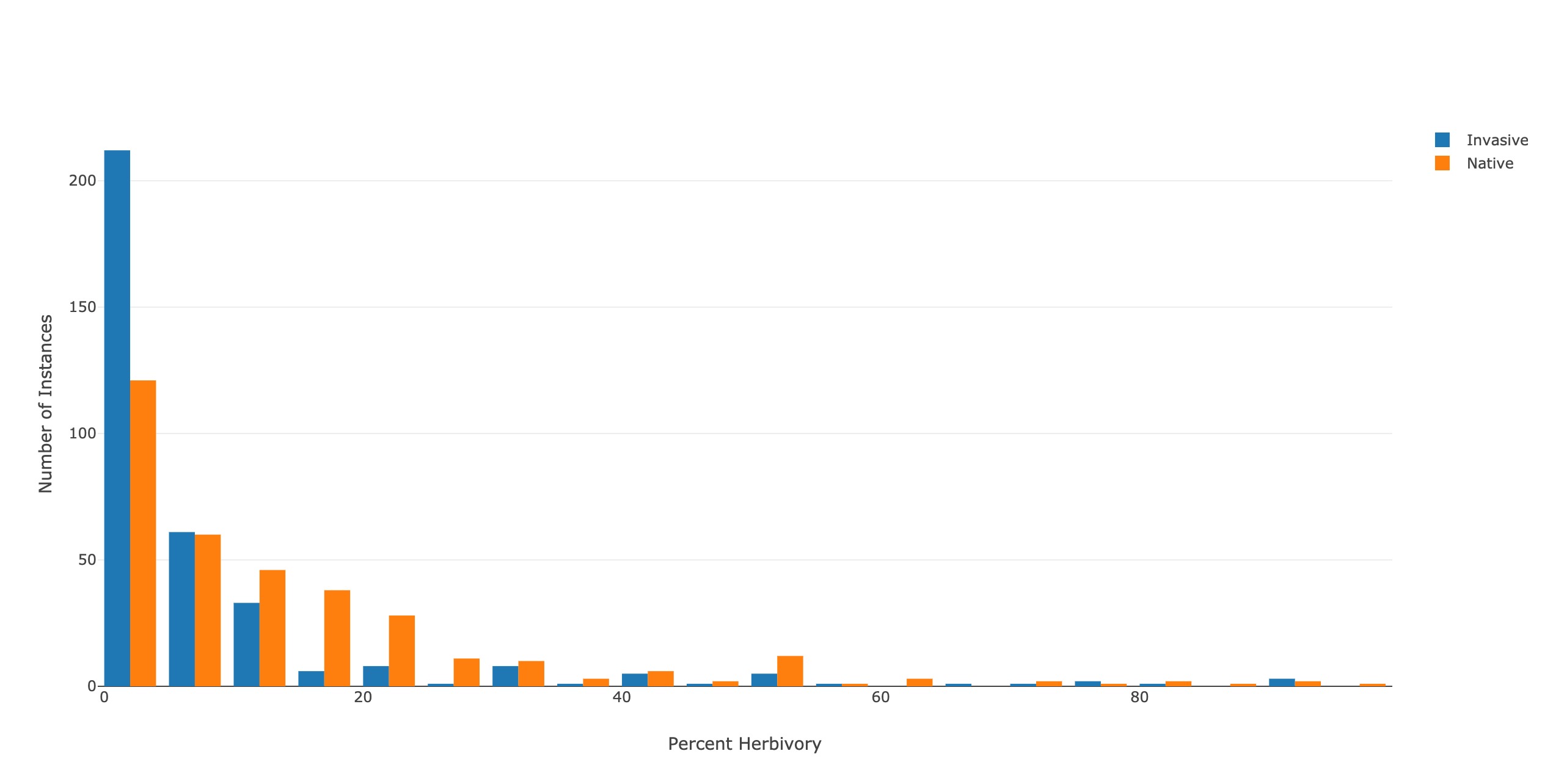}
         \caption{Query}
     \end{subfigure}
     \hfill
     \begin{subfigure}[b]{0.33\linewidth}
         \centering
         \includegraphics[width=\linewidth]{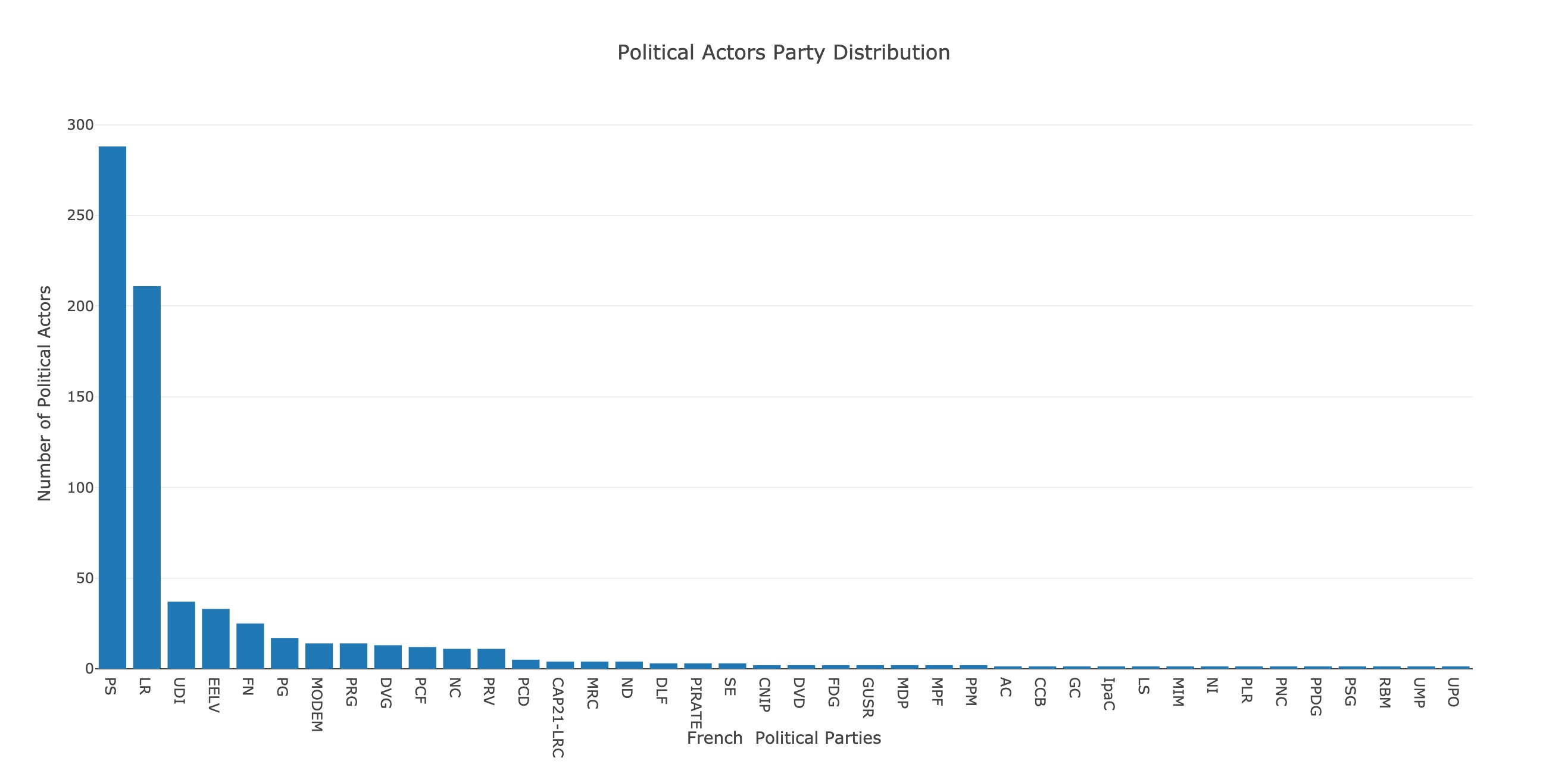}
         \caption{Visual Information~(CNN)}
     \end{subfigure}
     \hfill
     \begin{subfigure}[b]{0.33\linewidth}
         \centering
         \includegraphics[width=\linewidth]{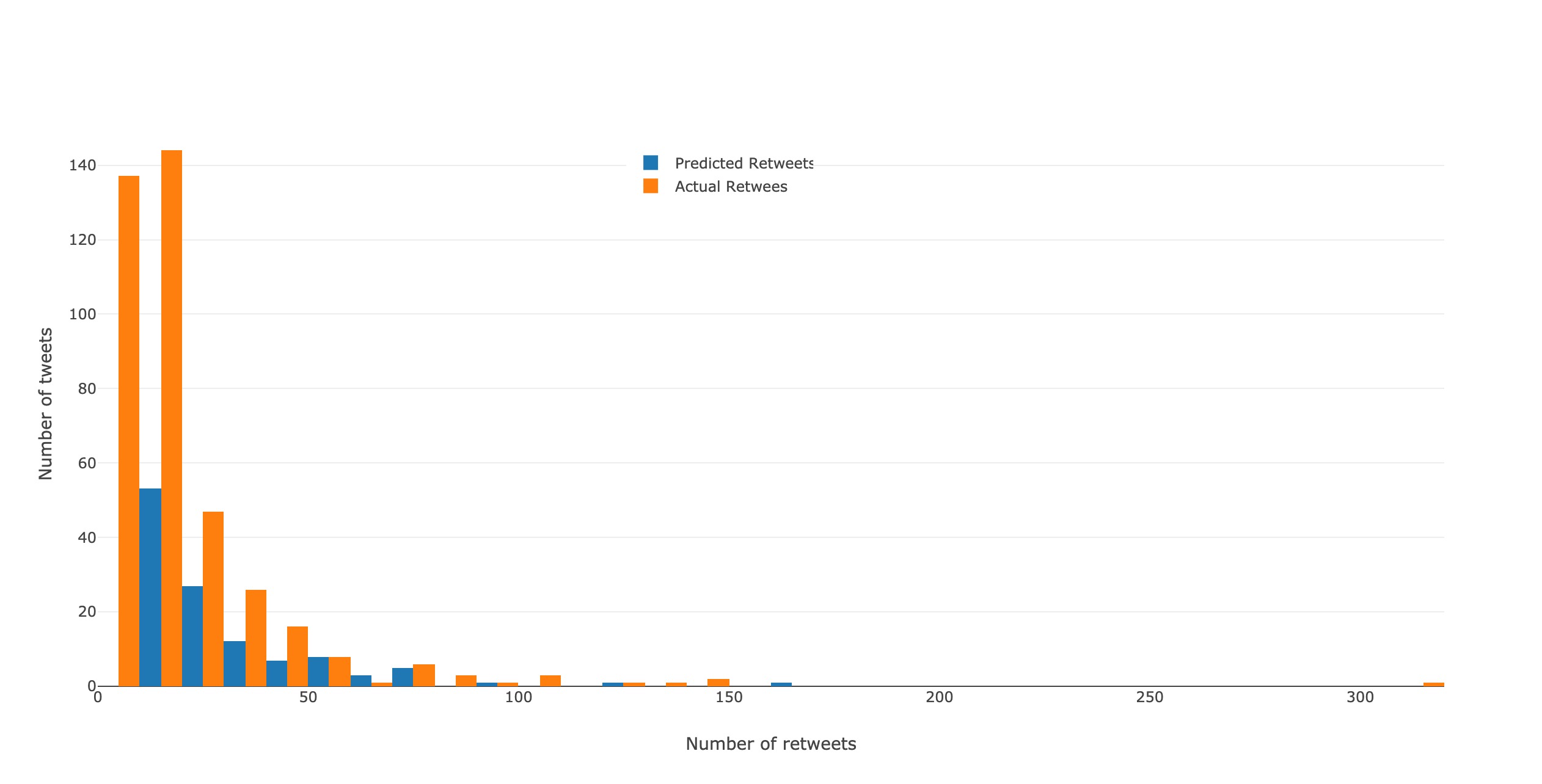}
         \caption{Structural and Visual Information}
     \end{subfigure}
        \caption{This figure shows (a) a sample query and the top-1 visualizations retrieved by (b) only considering visual information using CNN and (c) considering both structural and visual information. This example shows that structural information is essential in performing similar visualization retrieval.}
        \Description[]{This figure shows (a) a sample query and the top-1 visualizations retrieved by (b) only considering visual information using CNN and (c) considering both structural and visual information. This example shows that structural information is essential in performing similar visualization retrieval.}
        \label{fig:intro}
\end{figure*}

\revise{\italic{\textit{Structural information}} of visualizations mainly describes the spatial and hierarchical relationship between elements, such as
the position, grouping and hierarchy of the basic visual elements (e.g., \textit{<rect>} and \textit{<path>}).}
Compared with the commonly used \italic{\textit{visual information}} (i.e., the visual features to describe the appearance of visualizations) of visualizations, structural information enables a unique perspective to characterize the appearance of visualizations at the level of visual elements instead of
pixels.
It provides an accurate description of how different visual elements are organized in visualizations.
For example, as shown in Figure~\ref{fig:intro}, \revise{a grouped bar chart with two groups of bars (Figure~\ref{fig:intro}(a)) and a bar chart with only one group of bars} (Figure~\ref{fig:intro}(b)) seem to show the same trend
and are regarded as similar charts, if only the visual information is considered by using a computer vision-based method (e.g., convolutional neural network~(CNN) models).
However, the grouped bar chart actually shows how two sets of data are compared and it should not be treated as a similar visualization as the
\revise{bar chart with a single group of bars}.
Instead, another grouped bar chart (Figure~\ref{fig:intro}(c)) with both similar \revise{structure and appearance}
should be regarded similar to the query bar chart (Figure~\ref{fig:intro}(a)).  
From the example above, it is obvious that structural information plays an important role in characterizing the perceptual similarity between visualizations.
However, it still remains unclear what kind of structure-based features can be extracted and how these structure-based features can be leveraged to facilitate similar visualization retrieval.

In this paper, we aim to fill the research gap by leveraging both structural and visual information to accurately evaluate the perceptual similarity between visualizations.
We first conducted a preliminary study to better understand users' criteria on assessing the perceptual similarity of visualizations and identified the three most important criteria, i.e., the type of a visualization, the number of visual elements and the overall trend of visualized data.
Building upon these results, we propose to transform SVG-based visualizations to graphs and bitmap images that reflect the structure and the appearance of visualizations, respectively.
Then we utilize contrastive representation learning
to comprehensively delineate structural and visual information in a visualization with embedding vectors.
\italic{\textit{Contrastive representation learning}} is a type of self-supervised learning method and can minimize the distance between similar samples and maximize the distances between diverse samples~\cite{jaiswal2020contrastive}.
With contrastive learning, we avoid manually labeling the similarity between different visualizations, enabling us to easily generalize our approach to various visualizations.
Finally, we gain an embedding vector for each visualization that characterizes its structural and visual information and is used for retrieving similar visualizations.
\revise{Using the VizML corpus~\cite{hu2019vizml},} we extensively evaluate our approach through a crowdsourced user study, multiple case studies and quantitative comparisons. The results demonstrate the effectiveness of our approach.

The major contributions of this paper are summarized as follows:
\begin{itemize}
    \item We present a novel structure-aware approach to characterize the perceptual similarity between visualizations \revise{through embedding vectors}, which enables effective similar visualization retrieval.
    \item We conduct extensive evaluations including a crowdsourced user study with 50 participants, multiple case studies and quantitative comparisons with existing visualization retrieval methods. The results verify the effectiveness of our structure-aware visualization retrieval approach.
    \item We summarize the lessons we learned during exploring the usage of structural information in visualization retrieval. 
\end{itemize}
\section{Related Work}
The related work of this study can be categorized into three parts: retrieval of visualizations, visualization similarity estimation and visualization storage formats.

\subsection{Visualization Retrieval}
\revise{Visualization retrieval has attracted researchers' interests in recent years along with the increasing number of visualizations.
Based on the type of queries, the methods for retrieving visualizations can be mainly categorized into two classes~\cite{stitz2018knowledgepearls}, retrieval by definition and retrieval by example.}

\revise{\italic{\textit{Retrieval by definition}} refers to the methods where users can explicitly specify the criteria of retrieving visualizations using either programming language or natural language.
For example, Hoque and Agrawala~\cite{hoque2020searching} enable users to create a JSON-like specification to indicate their target characteristics of visualizations such as encoding types.
Some other prior studies~\cite{chen2015diagramflyer, li2015novel,siegel2016figureseer, stitz2018knowledgepearls} also provide users with tools to search for visualization using explicit queries.
Compared with retrieval by definition, \italic{\textit{retrieval by example}} provides an intuitive way for users to define the criteria of retrieving visualizations.
Users can use existing visualizations or sketches to search for other visualizations.
Several recent studies~\cite{saleh2015similarity, qian2020retrieve, ma2020scatternet} take example visualizations as inputs and return similar ones for data exploration or visualization re-use.
Zenvisage~\cite{siddiqui2016effortless} and ShapeSearch~\cite{siddiqui2020shapesearch} allow users to sketch their desired data pattern in visualizations.
Then they retrieve the data which matches the pattern from the database and visualize them to users.
In this line of research, one of the core problems is how to define the similarity between visualizations, which will be further discussed in Section~\ref{sec:vis_similarity}.}

\revise{Our structure-aware approach falls in the category of retrieval by example.
Our approach takes SVG-based visualizations as the input and then represents the visual and structural information of them as embedding vectors for similar visualization retrieval. }

\subsection{Visualization Similarity}\label{sec:vis_similarity}
\revise{Computing the similarity of visualizations benefits various downstream tasks such as assisting users in exploratory data analysis~\cite{zhao2020chartseer}, querying visualizations~\cite{ma2020scatternet} and generating visualization collections~\cite{opperman2021vizsnippets}.
Inspired by a previous study~\cite{ma2020scatternet}, prior methods on computing the similarity of visualizations can be roughly categorized into two classes: data similarity and perceptual similarity.}

\revise{The first class of methods solely focuses on the visualized data to delineate the similarity between visualizations.
Some representative studies in this class include SeeDB~\cite{siddiqui2016effortless},  ShapeSearch~\cite{siddiqui2020shapesearch} and VizCommender~\cite{oppermann2021vizcommender}.
SeeDB~\cite{siddiqui2016effortless} and ShapeSearch~\cite{siddiqui2020shapesearch} define the similarity between visualizations as the similarity of data distribution or trend.
These methods require that the raw data is available, which limits their application scenarios.
A possible way to mitigate the issue is to extract the raw data from visualizations and then calculate the data similarity. 
However, since the performance of existing data extraction methods (e.g.,~\cite{savva2011revision})  is not satisfactory~\cite{jung2017chartsense}, 
the inaccurately extracted data may further affect the results of retrieval.
}

\revise{Another class of methods focuses on the perceptual similarity of visualizations.
They extract visual features from visualizations and further utilize the distance between hand-crafted or learned feature vectors to characterize the similarity of visualizations. 
Hand-crafted features mainly refer to those features which are selected by the authors and can reflect certain characteristics of visualizations.
For example, 
prior studies proposed to use color histograms~\cite{saleh2015similarity} or histograms of gradients~\cite{opperman2021vizsnippets} to measure the perceptual similarity of infographics or visualization workbooks.
Due to the inefficiency and complexity of selecting hand-crafted features, representations automatically learned by machine learning models have been applied recently.
For example, 
ChartSeer~\cite{zhao2020chartseer} proposed to use an autoencoder to extract the representations of visualizations from their specifications.
ScatterNet~\cite{ma2020scatternet} and ChartNavigator~\cite{zhang2021chartnavigator} introduced convolutional neural networks~(CNNs) on visualizations to learn their representations.}

\revise{This paper aims to propose a structure-aware approach for retrieving perceptually similar visualizations.
Compared with the retrieval methods based on data similarity~(e.g.,~\cite{siddiqui2016effortless, vartak2015seedb}), our approach does not require the existence of original data and thus extends the scope of inputs.
Differing from the existing approaches which compute the perceptual similarity\footnote{In the remaining part of this paper, ``similarity'' refers to perceptual similarity.} of visualizations (e.g.,~\cite{oppermann2021vizcommender, ma2020scatternet}), our approach considers
both the pixel-level visual appearance and
the relationship between visual elements.
Such a design allows our approach to better match the crowdsourced criteria of perceptual similarity, which will be introduced in Section~\ref{sec:lessons}.}

\subsection{Visualization Format}
Depending on the ultimate purposes, visualizations can be stored in various formats including graphics, programs and hybrid approaches~\cite{wu2021survey}. 

The \italic{\textit{graphics}}-based visualizations include
two common formats: raster graphics and vector graphics. 
Raster graphics~(i.e., bitmaps) are the most common approaches to store and share visualizations~\cite{satyanarayan2020critical} due to their high compatibility.
However, they are hardly editable and often lead to the loss of visualization-specific information such as the chart type and visual encodings~\cite{wu2021survey}.
Alternatively, vector graphics such as SVGs provide general users with the flexibility of modification and annotation and can preserve partial visualization-related information~\cite{wu2021survey,wu2020mobilevisfixer} such as the relationship between visual elements.
Previously, some studies have explored utilizing vector graphics to achieve visualization query by specification~\cite{hoque2020searching} and visualization type classification~\cite{battle2018beagle}. 
 
\italic{\textit{Programs}} are also commonly used to store visualizations such as D3~\cite{bostock2011d3}, Vega-Lite~\cite{satyanarayan2017vegalite} or Plotly.
They preserve the visualization-related information and raw data. However, when rendering them as visualizations, extra compilers are always required, which limits their compatibility and wide usage.
Recently, to combine the advantages of graphics and programs, several studies have also explored \italic{\textit{hybrid approaches}} by embedding programs into graphics, such as~\cite{raji2021dataless,zhang2021viscode,chartem2021fu}.

Bitmaps and programs are commonly used in previous studies to compute the similarity of visualizations~\cite{wang2021survey,ma2020scatternet, zhao2020chartseer}.
However, bitmaps suffer from the lack of visualization-specific information while the usage of programs limits the generalizability of related methods.
Thus, in our paper, we propose to utilize the structural information~(see Section~\ref{sec:structural_info}) in scalable vector graphics~(SVGs), which is widely used in spreading visualizations on the Internet due to its interactivity~\cite{battle2018beagle}.
We combine the structural information extracted from SVGs and the visual information extracted from bitmaps to achieve effective characterization of visualization similarity.
\section{Background}
In this section, we introduce the background of our research, including the structural information in SVGs~(Section~\ref{sec:structural_info}), contrastive learning~(Section~\ref{sec:contrastive}), which enables unsupervised representation learning for visualizations, and graph neural networks~(Section~\ref{sec:gnn}).

\subsection{Structural Information in SVGs}\label{sec:structural_info}
Scalable vector graphics~(SVGs) are files used to describe vector-based graphics using Extensible Markup Language~(XML)\footnote{\url{https://developer.mozilla.org/en-US/docs/Web/SVG}}.
They have been widely used in visualizations on the Internet~\cite{battle2018beagle}. 
Compared with bitmaps, SVGs can preserve more visualization-specific information such as the style of the visual elements~\cite{wu2021survey}.
The structural information in SVGs mainly includes the hierarchical and spatial relationship among elements and the properties of each element.
The hierarchical information of visual elements can reflect how they are inherited and grouped, for example, in Figure~\ref{fig:graph_construction}, the \textit{<path>}s of green bars are grouped under the same \textit{<g>} while the \textit{<path>}s of blue bars are grouped under another \textit{<g>}.
Such information of visual elements can further illustrate the usage of visual channels and reveal some information of the raw data like the number of data instances and the number of attributes.
The spatial relationship, which is extracted based on the positions of visual elements, can describe how the elements are placed and a rough trend of data.
The properties of each element in SVGs also encode rich information, for example, the type, style, and shape of the element.
The types of elements reflect the functionality of elements, for example, \textit{<g>} is used to group other elements and \textit{<text>} can be rendered as graphics containing text.
The styles of an element indicate how the elements are rendered. 
Some common styles include the color and stroke of an element.
Furthermore, the shape of some visual elements can also be obtained from the properties, for example, the attribute \textit{``d''} in \textit{<path>} defines the shape of a path.
SVGs also have some properties related to interactions~(e.g., \textit{onclick} and \textit{onmouseover}).

In this paper, we propose to utilize the structural information extracted from SVGs to enhance the retrieval of similar static visualizations. 
To be more specific, we mainly consider structural information to reflect the hierarchical and spatial relationship among elements.
Some side information such as the types and styles of elements is also utilized to distinguish different elements.
More details are further illustrated in Section~\ref{sec:info_extraction}.

\subsection{Contrastive Learning}\label{sec:contrastive}
\begin{figure}[h!]
    \centering
    \includegraphics[width=\linewidth]{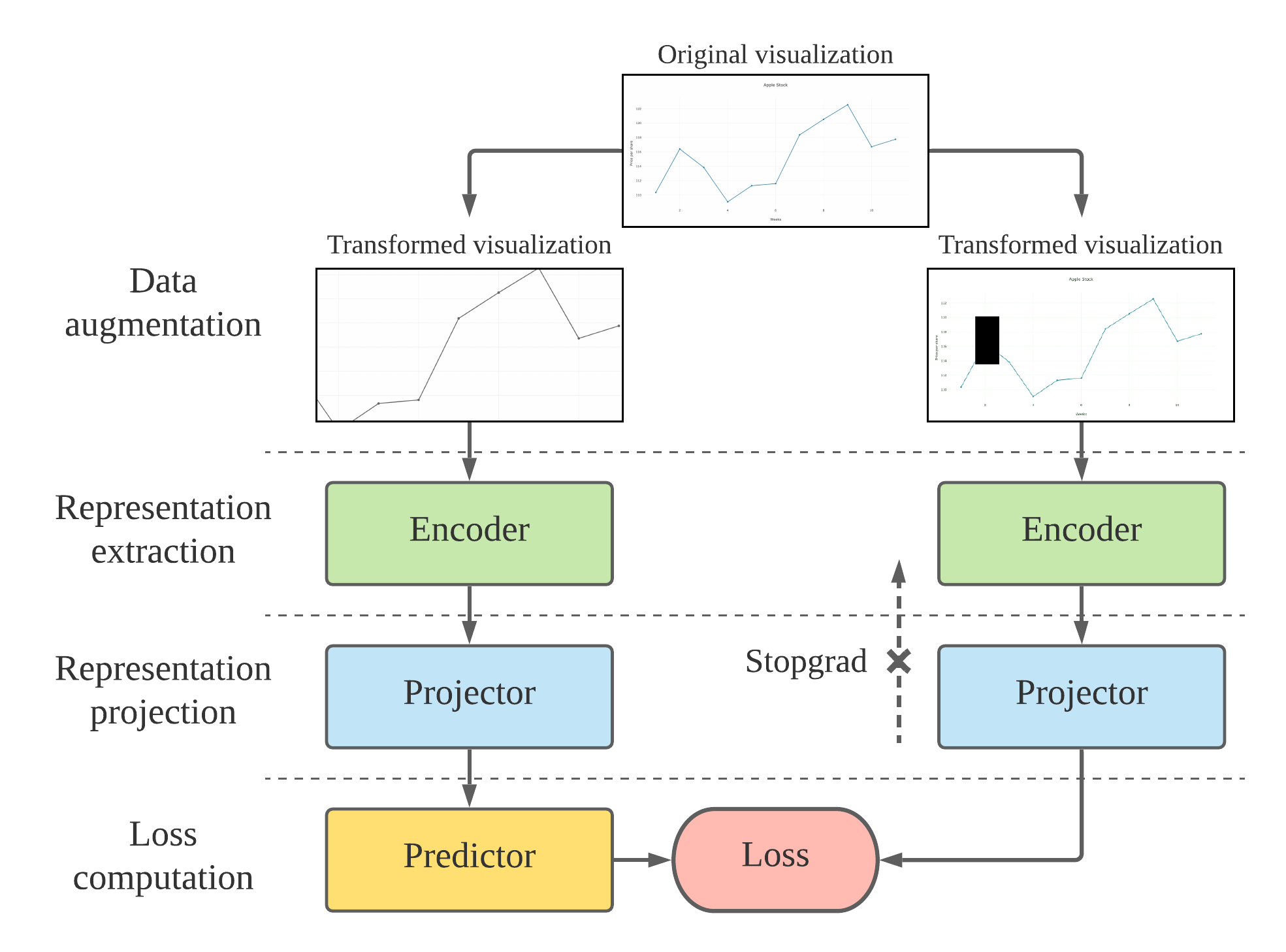}
    \caption{The basic structure of contrastive learning contains four major modules~\cite{chen2020simclr}: data augmentation, representation extraction, representation projection and loss computation. This figure shows SimSiam~\cite{chen2020simsiam} as an example.}
    \Description[]{The basic structure of contrastive learning contains four major modules: data augmentation, representation extraction, representation projection and loss computation. This figure shows SimSiam as an example.}
    \label{fig:simsiam}
\end{figure}

Supervised deep learning approaches always require a large number of samples with labels to train a model with satisfactory performance~\cite{jaiswal2020contrastive}.
However, corpora with high-quality labels are always hard to be obtained due to the high cost of human annotation.
Thus, to reduce the effort of manually labeling, self-supervised representation learning approaches, which are sometimes considered as a subset of unsupervised learning methods, have attracted researchers in various fields, for example, computer vision~\cite{chen2020simsiam, chen2020simclr}, user interface~(UI) design~\cite{li2021screen2vec} and visualization~\cite{tkachev2021s4}.
Contrastive learning is a representative approach of self-supervised learning.
The basic idea behind it is to train a model which can discriminate similar and dissimilar samples~\cite{jaiswal2020contrastive}.
As illustrated in Figure~\ref{fig:simsiam}, a common pipeline in contrastive learning approaches contains 4 steps: \italic{\textit{data augmentation, representation extraction, representation projection}} and \italic{\textit{contrastive loss computation}}~\cite{chen2020simclr}.
In the first step, a data sample will be randomly transformed~(e.g., distortion for images) and the transformed samples will be considered as similar~(positive) samples.
Then the transformed samples are encoded to embedding vectors by an encoder in the second stage.
The embedding vectors are further projected to a space where the loss is computed.
After training using the pipeline above, the encoder is used solely to extract representations of data samples and different projectors can be trained for various downstream tasks such as classification.

In our paper, we propose to use two CNN- and GNN-based contrastive learning models~\cite{chen2020simsiam, sun2019infograph} to generate the embedding vectors of visualizations' \revise{visual and structural} information, respectively.
Then the embedding vectors are concatenated and used for visualization retrieval.

\subsection{Graph Neural Networks}\label{sec:gnn}
\revise{Inspired by successful convolutional neural networks~(CNNs), graph neural networks~(GNNs) have been proposed to model the relationship among nodes in graphs.
The basic idea behind GNNs is to propagate the features of nodes through edges and then aggregate the information on nodes to capture node features and graph structures~\cite{zhou2020graph}.
The \italic{\textit{feature propagation}} and \italic{\textit{aggregation}} can be considered as a generalized convolutional filter on graphs.
GNNs have shown outstanding performance on graph-related tasks (e.g., node classification~\cite{schlichtkrull2018modeling} and graph classification~\cite{xu2018gin}) in various application domains (e.g., UI design~\cite{patil2021layoutgmn}, online education~\cite{li2020peer} and visualization~\cite{wang2019deepdrawing}).
As introduced in Section~\ref{sec:structural_info}, SVG elements are organized as trees that can also be regarded as graphs.
Thus, it is intuitive to apply GNNs to learn and represent the structural information in SVGs as embedding vectors.
Considering the advantages of contrastive learning (see Section~\ref{sec:contrastive}), GNN-based contrastive learning approaches are suitable for extracting the embedding vectors of graphs of SVG elements in our structure-aware approach.
Existing GNN-based contrastive learning approaches
are mainly applied to three types of tasks~\cite{wu2021self}, including node-level tasks such as node classification (e.g., DGI~\cite{velivckovic2018deep}), edge-level tasks such as link prediction (e.g., BiGi~\cite{cao2021bipartite}) and graph-level tasks such as graph classification (e.g., InfoGraph~\cite{sun2019infograph}).
Since we aim to represent the structural information in the graph of SVG elements with embedding vectors, InfoGraph~\cite{sun2019infograph}, one of the state-of-the-art methods for graph embedding, is applied in our approach. 
Section~\ref{sec:embedding_svg} will introduce more details of InfoGraph.
}

\section{Preliminary Study}
Before designing our structure-aware visualization retrieval approach, we conducted a preliminary study in which we collected the opinion of 54 visualization users on the criteria of perceptual similarity between visualizations. 
In this section, we introduce the procedure\footnote{The protocol of the preliminary study and the user study has been approved by the Institutional Review Board of our institution.} of the study and summarize the important criteria.

\subsection{Procedure}
Our preliminary study was conducted on Prolific\footnote{\url{https://www.prolific.co/}}, a widely used platform for recruiting research participants.
\revise{There are totally three parts in the study.
In the first part,}
the overall process was introduced to our participants prior to obtaining their consents to join the study.
Then, to verify that the participant has basic knowledge of visualizations, each participant was required to answer three simple visualization-related questions, for example, \textit{``what is the chart type of the given visualization?''}.
Only participants who correctly answered the three verification questions were allowed to join the study.
\revise{
No other criteria were used in the participant recruitment.}
\revise{In the second part of the study}, to encourage the participants to reflect on how they judge the similarity of visualizations, each participant was presented with five query visualizations and their retrieved \revise{top-5} similar visualizations by using visual information only.
The participants were asked to give each retrieved visualization a score ranging from 1~(the least similar) to 5~(the most similar).
After finishing the scoring, \revise{in the last part of the study}, we asked participants to write down their criteria of scoring the retrieved visualizations \revise{in a text box}.

After the study, we summarized the responses from participants. 
Since there may be ambiguity in understanding the criteria mentioned by participants, we first classified the major criteria into six major categories and two co-authors of this paper labeled all responses individually.
If the annotations were inconsistent on any response, we examined and discussed together to reach an agreement on these cases.

\begin{figure}[h!]
    \centering
    \includegraphics[width=\linewidth]{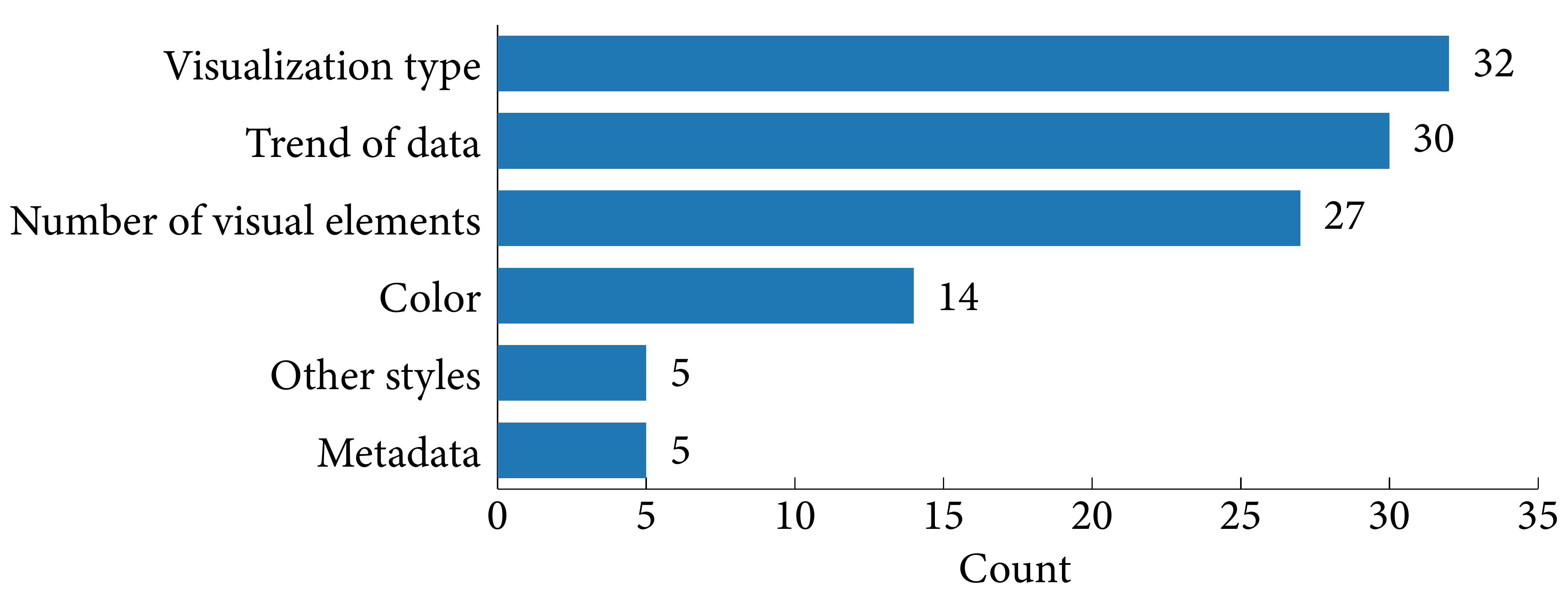}
    \caption{We categorized different criteria mentioned by participants into six classes and identified the three most important criteria based on their frequency. \italic{\textit{Other styles}} refer to the styles of visual elements other than colors, for example, the space between bars and the width of bars. \italic{\textit{Metadata}} refers to the meta-information of data in the visualizations, for example, the range and the type of data.}
    \Description[]{This bar chart shows the frequency of different criteria. Visualization type has been mentioned 32 times. Trend of data has been mentioned 30 times. Number of visual elements has been mentioned 27 times. Color has been mentioned 14 times. Other styles and metadata have been mentioned 5 times. Other styles refer to the styles of visual elements other than colors, for example, the space between bars and the width of bars. Metadata refers to the meta-information of data in the visualizations, for example, the range and the type of data.}
    \label{fig:pre_study}
\end{figure}

\subsection{Results}
The six major criteria and their frequency are shown in decreasing order in Figure~\ref{fig:pre_study}.
In the results, we can notice that there are three important criteria~(i.e., visualization type, the trend of data and the number of visual elements) 
with much higher frequency than other criteria.
The results of our preliminary study also align with a previous study~\cite{kim2021automated} well.
\revise{Specifically, the number of visual elements and the trend of data are also considered when measuring the difference between two visualizations in the previous research~\cite{kim2021automated}.}
Thus, the type of visualization, the trend of data and the number of visual elements are necessary to be considered explicitly in our approach when characterizing similarity of visualizations.

\section{Method}
\begin{figure}[h!]
    \centering
    \includegraphics[width=\linewidth]{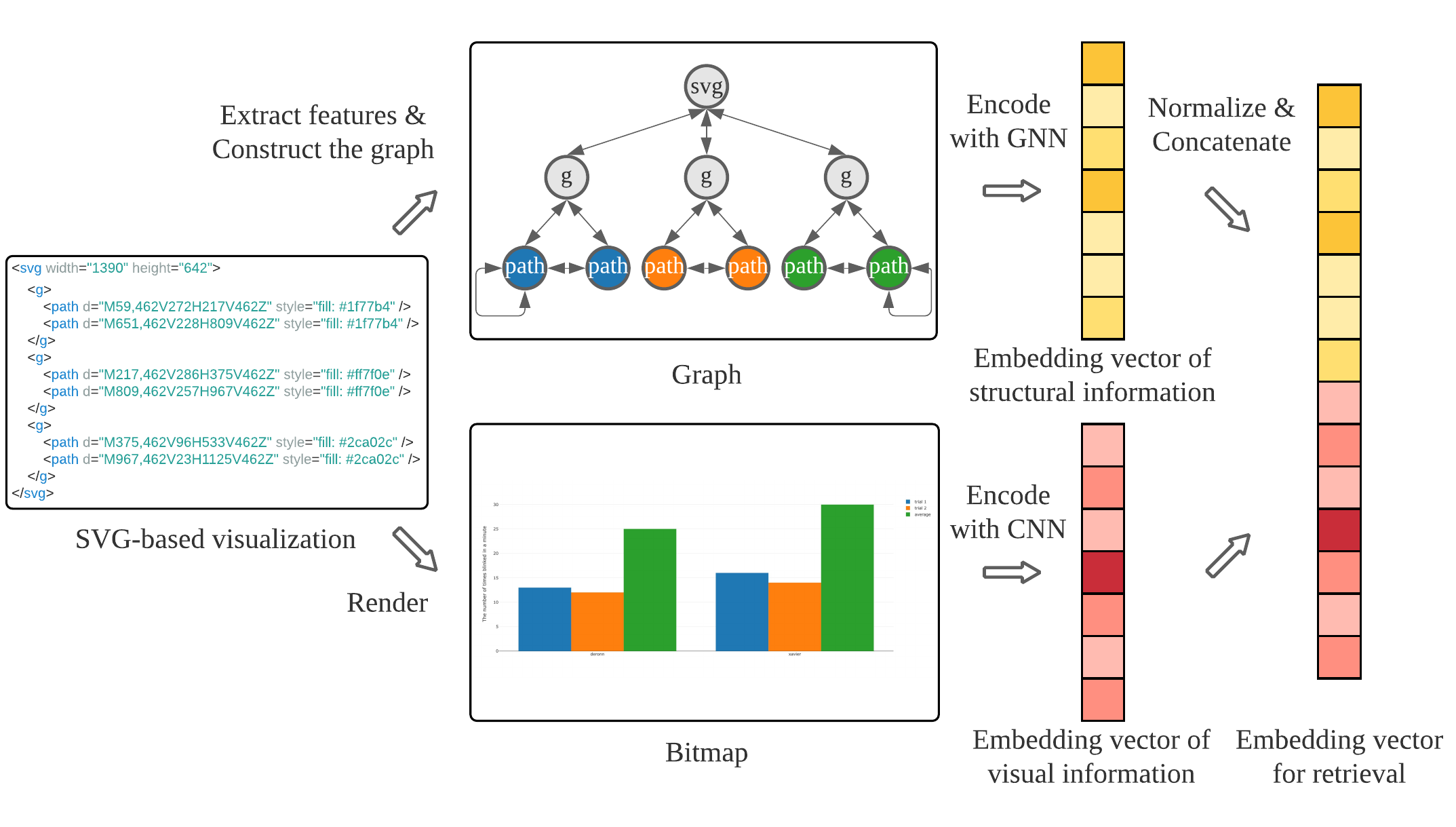}
    \caption{Our approach extracts both structural and visual information from visualizations first. Then the two types of information are encoded to embedding vectors separately. Finally, two embedding vectors are normalized and concatenated as the final representation of the visualization.}
    \Description[]{Our approach extracts both structural and visual information from visualizations first. Then the two types of information are encoded to embedding vectors separately. Finally, two embedding vectors are normalized and concatenated as the final representation of the visualization.}
    \label{fig:overview}
\end{figure}

In this section, the method of our structure-aware visualization retrieval is introduced.
An overview is shown in Figure~\ref{fig:overview}.
To extract and represent the structural information in a visualization, we first construct a graph of visual elements with features and then apply a GNN encoder to generate the embedding vector of it~(Section~\ref{sec:embedding_svg}).
Then we also render the visualization to a bitmap and use a CNN model to encode the visual information as an embedding vector as well~(Section~\ref{sec:embedding_bitmap}).
Here we applied contrastive representation learning to train both CNN and GNN encoders since it can eliminate human efforts on data annotation.
Finally, we normalize and concatenate the embedding vectors of structural and visual information for similar visualization retrieval~(Section~\ref{sec:vis_retrieval}).

\begin{figure*}[h!]
    \centering
    \includegraphics[width=\linewidth]{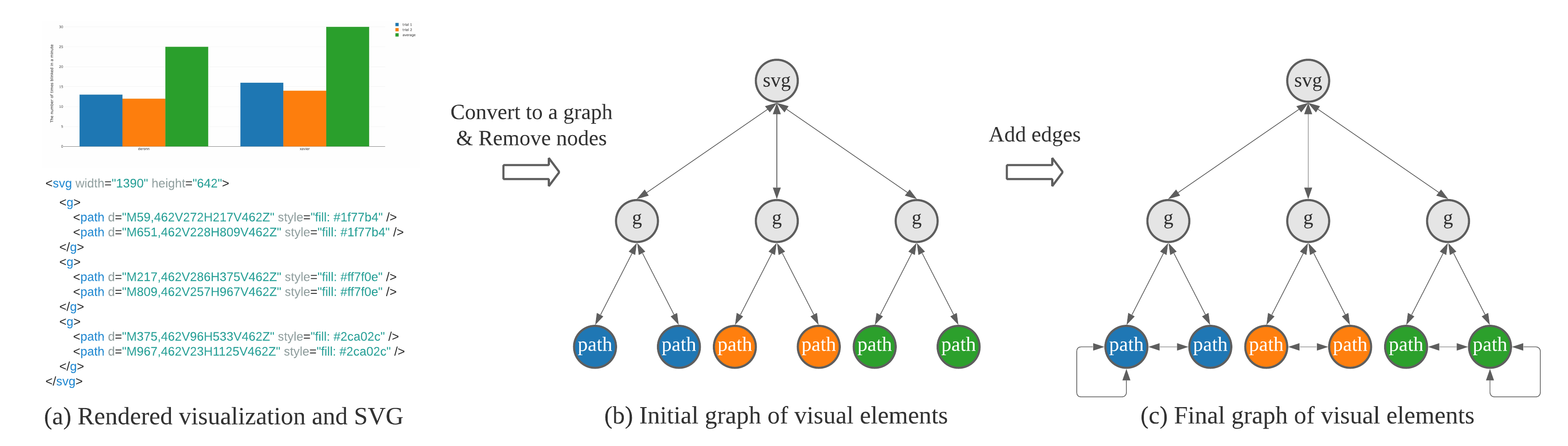}
    \vspace{-1em}
    \caption{This figure illustrates how we transform an SVG to a graph of visual elements. In the SVG, each bar can be either represented by a \textit{<path>} or \textit{<rect>}. Due to the limited space, we only show partial SVG and the corresponding subgraph. Each node represents an element in the SVG. The colors of nodes in the graph indicate the corresponding bars in the visualization. In (c), all leaf nodes are with self-loop edges, which are not all shown in this figure. }
    \Description[]{This figure illustrates how we transform an SVG to a graph of visual elements. In the SVG, each bar can be either represented by a <path> or <rect>. Due to the limited space, we only show partial SVG and the corresponding subgraph. Each node represents an element in the SVG. The colors of nodes in the graph indicate the corresponding bars in the visualization. In (c), all leaf nodes are with self-loop edges, which are not all shown in this figure. }
    \label{fig:graph_construction}
\end{figure*}

\subsection{Representation Learning of Structural Information}\label{sec:embedding_svg}\label{sec:info_extraction}
As introduced before, structural information in SVGs can reflect the hierarchical and spatial relationship between visual elements explicitly.
To utilize the structural information, 
we first extract visual element-level features and construct a graph of visual elements.
Then, we apply a GNN-based graph contrastive learning method to generate the embedding vector of the structural information.

\textbf{Feature Extraction.} In the first step, we aim to extract features to describe the characteristics of elements in SVGs. 
These features are designed to reflect the types, styles, shapes and positions of elements.
To make our approach simple and generalizable, we only extract basic features inspired by Beagle~\cite{battle2018beagle}.
The \italic{\textit{types}} of elements can reflect their functionality in an SVG.
We represent the types of elements with one-hot encoding.
The \italic{\textit{style}} features of elements mainly include the color, the stroke width and the opacity of the element, which are the most common styles of elements.
\revise{
According to the results of our preliminary study, we also consider the styles of SVG elements.
Though the styles of visualizations such as colors are not the most frequent criteria when deciding the similarity between visualizations, it is still considered by some participants.
Thus, we take the commonly used styles into consideration and ignore those infrequent ones such as the stroke style.}

How to define \italic{\textit{shape}} features for different types of visual elements is challenging due to their characteristics.
For some visible visual elements like bars in bar charts and scatters in scatter plots, we are able to describe each visual element with the area, center, height and width of its bounding box.
However, it is not enough to describe lines in line charts with these simple features. 
Two lines with the same bounding box can represent totally different trends of data.
To deal with this issue, we further introduce two features to describe the shape of lines in line charts: the number of vertices in the line~\cite{battle2018beagle} and the trend of the line.
Inspired by a prior study~\cite{kim2021automated}, LOESS regression~\cite{cleveland1979loess} is applied to model the rough trend of a line based on vertices.
Since LOESS is a non-parametric regression approach, we are not able to extract a fixed number of features to describe the trend.
Thus, we use the predicted values of LOESS on five evenly sampled vertices as features of the trend.
Since \textit{<text>} can hardly be described by the features above, we further add the length of the text as a feature.
Furthermore, the relationship between positions of elements within the same group is also necessary to reflect the overall trend in the visualizations.
Thus, we sort the visual elements according to their positions on the horizontal and vertical axes and introduce the differences in positions as features of each element as well.
Finally, for elements without certain features~(e.g., \textit{<g>} does not have specified width and height), we fill zeros as the placeholders.
All features are also scaled according to their ranges. For example, positions are scaled using the height and width of the entire SVG.

\textbf{Graph Construction.}
After extracting the features for individual elements in visualizations, we construct graphs of visual elements based on the tree structure of elements in SVGs.
Such graphs of visual elements are used to represent the hierarchical and spatial relationships between different visual elements in visualizations.
Since the graph is the input to a graph neural network~(GNN) in the latter stage, we further process them to facilitate the propagation of features between nodes.
An overview of graph construction is shown in Figure~\ref{fig:graph_construction}.

First, the tree of elements in an SVG is transformed to a bidirectional graph where elements and hierarchical relationships are treated as nodes and edges, respectively.
Then, we remove the visual elements which \revise{serve as references such as} the legend and grid lines.
This step is conducted to reduce the noise in the graph of visual elements.
In the next step, we remove the \textit{<g>} elements that connect with one or two other elements.
The rationale behind this operation is that \textit{<g>} is used to group other elements and thus not meaningful if it has zero or one child.
Removing these \textit{<g>} elements can reduce unnecessary nodes and edges in the graph.
Finally, we add two types of edges to augment the graph: the self-loop edges and the edges between neighbor elements~(see Figure~\ref{fig:graph_construction}(c)).
The self-loop edges are added to preserve each element's own features during the feature propagation.
The edges between neighbor elements are designed to reflect the spatial relationship between elements by propagating their spatial features.
Here neighbor elements are defined as the elements which are next to each other after sorting according to their positions as described previously.

\textbf{Contrastive Learning Structure.} To represent the structural information as an embedding vector, we utilize InfoGraph~\cite{sun2019infograph}, a state-of-the-art graph contrastive representation learning model.
The core idea of InfoGraph is to maximize the mutual information between the embedding vector of the whole graph and the embedding vectors of substructures~(e.g., individual or a group of nodes) in the same graph.
To be more specific, it takes pairs of graphs as the input of a GNN encoder.
Then it optimizes the encoder by maximizing the mutual information between the embedding vector of one graph and the embedding vector of a subgraph inside, and minimizing the mutual information between the embedding vectors of a graph and the subgraph in another one.
Finally, it \revise{uses} the encoder to generate the embedding vectors of all nodes and \revise{aggregates} them as the representation of the graph.

In our method, we train InfoGraph to generate the embedding vectors of graphs of visual elements, which encode the structural information in SVGs.

\subsection{Representation Learning of Visual Information}\label{sec:embedding_bitmap}
In the previous section, we introduce how we extract the representation of structural information hidden in SVGs with a state-of-the-art graph contrastive learning method.
Though the structural information can \revise{partially reflect the appearance} of a visualization, for example, the size and position of a visual element, it still lacks the ability to comprehensively describe the appearance of visualizations.
Thus, inspired by previous studies~\cite{zhang2021chartnavigator, ma2020scatternet} which utilize convolutional neural networks~(CNNs) to extract the visual features of visualizations, our approach utilizes contrastive learning to generate the representation of visualizations using bitmaps as input. 
In the rest of this section, we briefly introduce the structure of the contrastive learning structure we used and then illustrate how we adapt it to extract visual information from visualizations.

\textbf{Contrastive Learning Structure.} 
In our method, SimSiam~\cite{chen2020simsiam} is applied to generate the representation of visual information.
Compared with other state-of-the-art contrastive learning approaches such as SimCLR~\cite{chen2020simclr} and BYOL~\cite{grill2020byol}, SimSiam has a similar performance with less demand for computational resources.
It utilizes a structure based on Siamese networks~(see Figure~\ref{fig:simsiam}).
Two randomly transformed samples~(denoted as $s_1$ and $s_2$) of the original sample are fed into the same encoder which is followed by a projection head.
After projection, both samples are transformed to embedding vectors~(denoted as $\mathbf{v_1}$ and $\mathbf{v_2}$, respectively), which are further fed into a prediction head with a bottleneck structure.
The outputs of the prediction head are denoted as $\mathbf{p_1}$ and $\mathbf{p_2}$ correspondingly.
Then the loss is calculated based on the negative cosine similarity with a stop-gradient operation as $Loss = -1/2*(cos(\mathbf{v_1}, stopgrad(\mathbf{p_2}))+cos(\mathbf{v_2}, stopgrad(\mathbf{p_1})))$.
Here the stop-gradient operation controls how the model is optimized with gradients and proved to be beneficial~\cite{chen2020simsiam}.

\begin{figure}[h!]
     \centering
     \begin{subfigure}[b]{0.48\linewidth}
         \centering
         \includegraphics[width=0.9\linewidth]{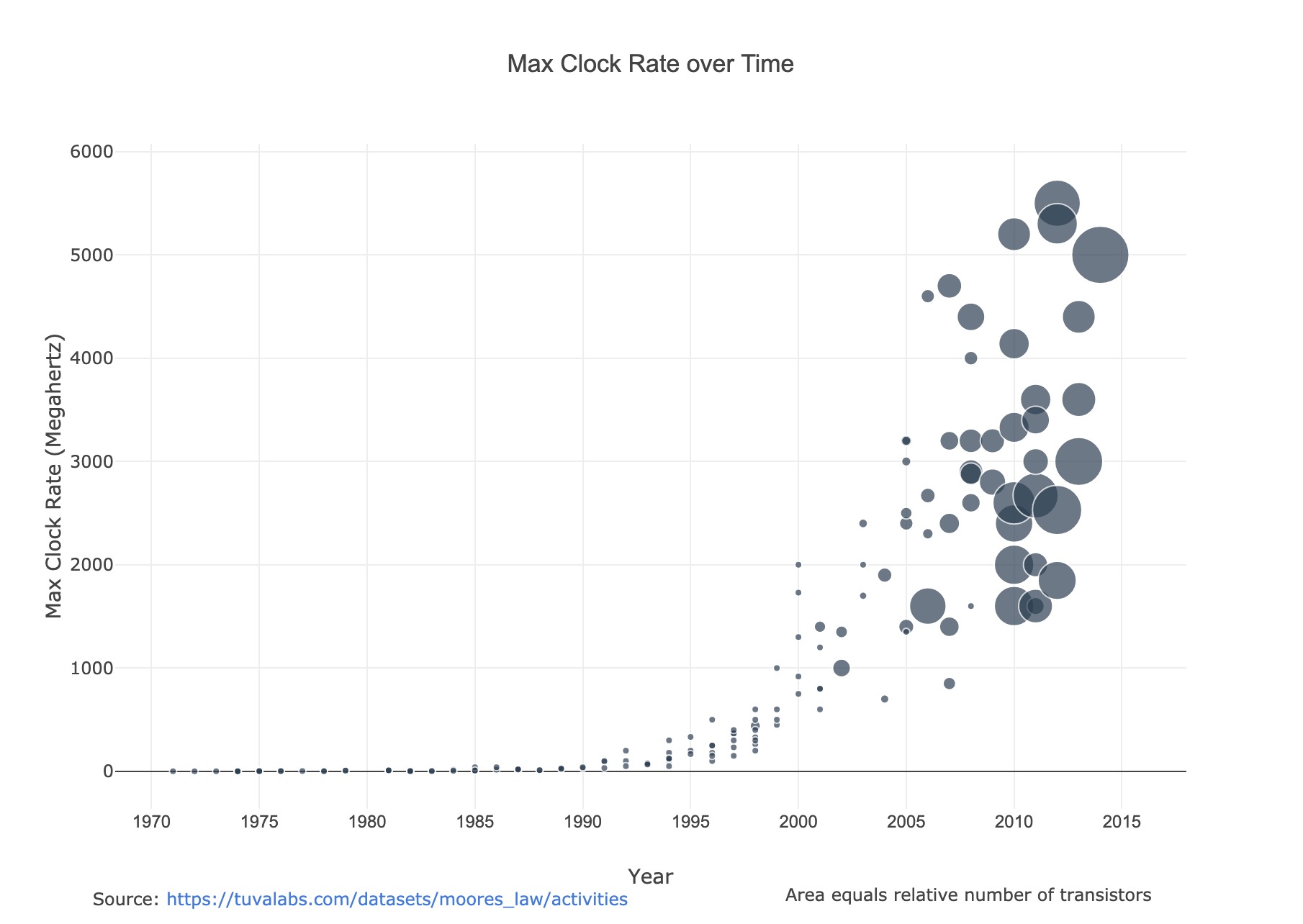}
         \caption{Original}
     \end{subfigure}
     \hfill
     \begin{subfigure}[b]{0.48\linewidth}
         \centering
         \includegraphics[width=0.9\linewidth]{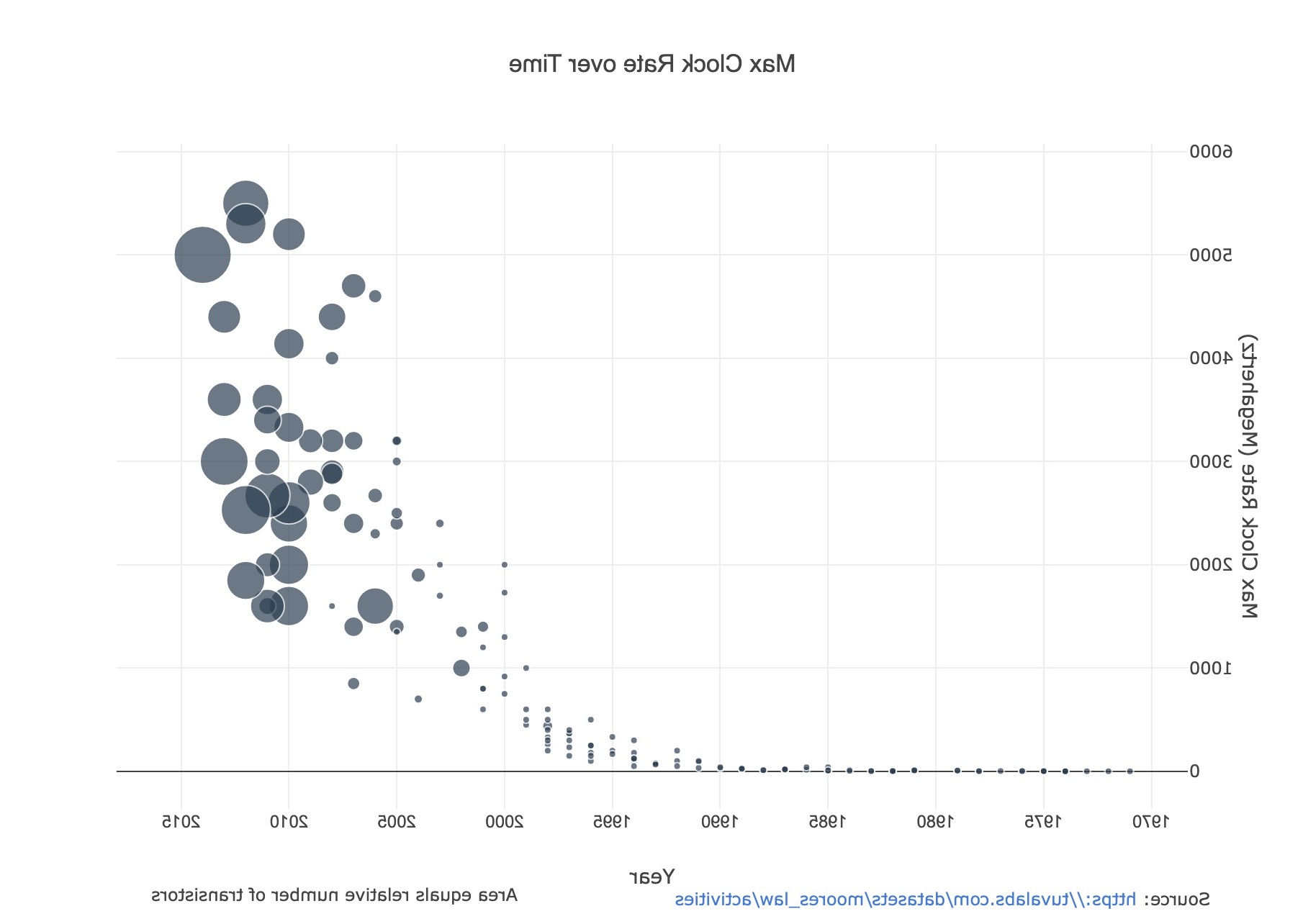}
         \caption{Flipped}
     \end{subfigure}
        \caption{The flipped visualization is dissimilar to the original one due to the change of overall trend. Thus, flipping is invalid for generating similar samples.}
        \Description[]{This figure shows an image of a scatter plot and its flipped result. The scatters in the original visualization have a growing trend while the scatters in the flipped one have a decreasing trend. The flipped visualization is dissimilar to the original one due to the change of overall trend. Thus, flipping is invalid for generating similar samples.}
        \label{fig:invalid_operation}
\end{figure}

\textbf{Pre-processing and Data Augmentation.} 
The original SimSiam is designed for general images.
To adapt it for visualizations, we modify the pre-processing of input bitmaps and data augmentation.
First, since the input of the encoder has to be a square bitmap, we need to resize or pad visualizations to be square.
Resizing input images is used in SimSiam~\cite{chen2020simsiam}.
However, resizing is not appropriate for visualizations since a resized visualization can alter the trend of the original one~\cite{kim2021automated}, which may affect the judgment on the similarity of visualizations.
Thus, we pad the original bitmap to be a squared one, which can better preserve the original trend of data.
The next step is to apply data augmentation techniques to generate transformed data instances as mentioned in Section~\ref{sec:contrastive}.
Chen~\textit{et al.}~\cite{chen2020simclr} have summarized eight common approaches used for image augmentation, including cropping and resizing, cutting out, flipping, rotating, blurring, applying noise or filter and distorting the colors.
However, not all of them are suitable for visualizations.
For example, flipping the visualization can generate a visualization with a different trend of data as shown in Figure~\ref{fig:invalid_operation}.
Thus, in the training of our CNN model for visualizations, we select some of the data augmentation techniques based on those used in SimSiam~\cite{chen2020simsiam}.
First, cropping and resizing, and partially cutting out are applied inspired by the closure principle in Gestalt principles of perception, which states that people have the ability to fill in the blanks and make the object in the image complete.
Also, the color distortion is randomly applied to some visualizations by adding color jitter or transforming the bitmap to a greyscale one.
The rationale behind this is that color is not always used to encode information in visualizations and is considered not the most important criteria according to our preliminary study~(see Figure~\ref{fig:pre_study}).
In Figure~\ref{fig:simsiam}, we present two transformed visualizations in which the left one is cropped and converted to a greyscale one while the right one is cut out and added color jitter~(padding is not shown in the figure).

After generating the transformed samples, we use the original settings of SimSiam to train the encoder.
At the end of the training, other parts of the model are disposed and the encoder is kept to generate the embedding vector of visual information.

\subsection{Visualization Retrieval}\label{sec:vis_retrieval}
In Sections~\ref{sec:embedding_svg} and~\ref{sec:embedding_bitmap}, we introduce how to represent both structural and visual information using low-dimensional embedding vectors.
Then, with the two embedding vectors of a visualization, we perform similar visualization retrieval.

In our approach, we normalize and concatenate both the embedding vectors into one vector, which is the final representation of a visualization.
Initially, we tried to apply multi-modal autoencoders~\cite{ngiam2011multimodal,feng2014cross} to learn the joint embedding vector of both structural and visual information.
However, the performance of the learned embedding vector using quantitative metrics in Section~\ref{sec:quantitative} is not better than simple concatenation.
We suspect the reason is that encoding two vectors into a single vector is a lossy compression, which introduces extra noise.
Thus, to avoid extra training and additional noise, we use concatenation as our approach to generate the final embedding vector of a visualization.
After obtaining the embedding vectors of visualizations, the cosine similarity between the embedding vectors of two visualizations is used to measure the distance between them.
When a visualization is selected for query, we calculate its similarity to other visualizations and rank the similarity scores to retrieve the most similar ones.

\section{Evaluation}
By using the large-scale SVG-bitmap visualization corpus~(Section~\ref{sec:corpus}), we conducted extensive evaluations to assess our structure-aware visualization retrieval approach including quantitative comparisons~(Section~\ref{sec:quantitative}), a user study~(Section~\ref{sec:user_study}) and case studies~(Section~\ref{sec:case_study}). 
We also introduce the model settings used in evaluations~(Section~\ref{sec:model_setting}).
The results verify the effectiveness of our structure-aware visualization retrieval approach.
\subsection{Corpus}\label{sec:corpus}
Since our approach is a deep learning-based method~(Sections~\ref{sec:embedding_svg} and \ref{sec:embedding_bitmap}), we need a relatively large-scale corpus to train and test it.
Thus, by using the links to visualizations provided in VizML~\cite{hu2019vizml}, we built a crawler to collect the SVG-bitmap visualization pairs \revise{from Plotly Chart Studio\footnote{\url{https://chart-studio.plotly.com/}}}.
Following previous practices using the VizML corpus~\cite{hu2019vizml, li2021kg4vis}, we also kept one visualization per user.
Also, we removed those pairs with invalid bitmaps or SVGs, for example, empty visualizations or incomplete SVGs.
Finally, 51,037 SVG-bitmap visualization pairs were collected.
The corpus contains five types of visualizations: bar charts, box plots, histograms, line charts and scatter plots.
Since the difference of histograms between bar charts mainly lies in the data transformation, which is beyond our scope, they are categorized into the same class in the following evaluations. 
A detailed distribution of visualization types is shown in Table~\ref{table:corpus}.
In our evaluations, 
we followed the practice in Screen2Vec~\cite{li2021screen2vec} and randomly sampled 90\% of pairs of each visualization type as the training set and the rest was used for testing.

\begin{table}[h!]
\caption{Statistics of the SVG-bitmap visualization corpus.}
\Description[]{Statistics of the SVG-bitmap visualization corpus.}
\centering
\small
\setlength{\aboverulesep}{0.5pt}
\setlength{\belowrulesep}{0.5pt}
\begin{tabular}{l|p{2cm}|p{2cm}}
\toprule
\multicolumn{2}{l|}{Number of SVG-bitmap pairs} & 51,037  \\ \hline
\multirow{4}{*}{Number of each visualization type} & Bar/Histogram &  12,608\\ \cline{2-3}
& Box & 3,269 \\ \cline{2-3}
& Line & 15,488 \\ \cline{2-3}
& Scatter & 19,672 \\ \bottomrule
\end{tabular}
\label{table:corpus}
\end{table}

\newcommand{\hog}{\textit{V-HOG}}
\newcommand{\cnn}{\textit{V-CNN}}
\newcommand{\svg}{\textit{S-GNN}}
\newcommand{\ours}{\textit{S\&V-Fusion}}

\subsection{Model Settings}\label{sec:model_setting}
We compared the retrieved visualizations using four methods to demonstrate the effectiveness of our structure-aware approach:
\begin{itemize}
    \item \textbf{Visual Information by Histogram of Oriented Gradients~(HOG).} HOG~\cite{dalal2005hog} is one of the most widely used feature descriptors of images and has been applied to compute the similarity between visualizations in a recent study~\cite{opperman2021vizsnippets}.
    \item \textbf{Visual Information by CNN.} CNNs have been widely applied to extracting the visual information of visualizations and computing the similarity~\cite{opperman2021vizsnippets, zhang2021chartnavigator, ma2020scatternet}. 
    Following the method described in Section~\ref{sec:embedding_bitmap}, we extract the visual information using SimSiam~\cite{chen2020simsiam} with ResNet-50~\cite{he2015resnet} as the encoder. 
    Empirically, we set the training epochs as 200, the batch size as 128 and the learning rate as 0.025. The embedding dimension is set as 512. 
    We did not use any pre-trained model as suggested by Haehn~\textit{et al.}~\cite{haehn2019evaluating}.
    \item \textbf{Structural Information by GNN.} 
    As mentioned in Section~\ref{sec:embedding_svg}, we use InfoGraph~\cite{sun2019infograph} with Graph Isomorphism Network~(GIN)~\cite{xu2018gin} as the encoder to generate the embedding vectors of structural information. 
    Based on the original experiment settings, we set the training epoch as 40, the batch size as 128 and the learning rate as 0.001. 
    Since the radii of graphs in our corpus are mostly lower than or equal to 2, we set the layers of the encoder as 2. The embedding dimension is 512 as well.
    \item \textbf{Structural and Visual Information Fusion.} This is our structure-aware approach which jointly considers structural and visual information by normalizing and concatenating the embedding vectors generated by CNN and GNN as mentioned in Section~\ref{sec:vis_retrieval}.
    \end{itemize}

After extracting the embedding vectors of visualizations using the four approaches above, we calculate the similarity score between visualizations as mentioned in Section~\ref{sec:vis_retrieval}. 
In the following sections, we use \textbf{\hog}, \textbf{\cnn}, \textbf{\svg} and \textbf{\ours} to simplify the names of the four approaches above.

\subsection{Quantitative Evaluation}\label{sec:quantitative}
We first conducted a series of quantitative evaluations to assess the effectiveness of our approach in comparison with baselines.

\textbf{Metrics.} As described in Section~\ref{sec:corpus}, there are several types of visualizations in our corpus.
Thus, we need to use some metrics which are unified across different types of visualizations.
Two simple metrics are utilized to approximately probe the performance of different approaches based on the criteria identified in our preliminary study.
The first metric is to measure the \italic{\textit{visualization type consistency}} between a query visualization and the retrieved visualizations.
To be more specific, we count the number of retrieved visualizations that are of the same type as the original visualization in a top-k query.
Then the number of such visualizations is divided by $k$ to obtain the \italic{\textit{type-consistent rate}} of each query visualization.
Finally, the average and the standard deviation of \underline{T}ype-\underline{C}onsistent \underline{R}ates of all visualizations in the test set are computed and  denoted as average type-consistent rate~($TCR_{ave}$) and standard deviation of type-consistent rates~($TCR_{std}$).

The second metric is to measure the \italic{\textit{difference of the numbers of visible visual elements}} between query and retrieved visualizations.
Here the visible visual elements refer to the elements that are the leaf nodes in our graph.
Since the non-leaf elements~(i.e., \textit{<g>} and \textit{<svg>}) are mainly used to group or contain other elements and are not directly visible, they cannot be observed by visual information-based models~(i.e., \cnn~and \hog).
To make the metric fair to all methods, we only count the visible visual elements.
Examples of visible visual elements are shown as colored nodes in Figure~\ref{fig:graph_construction}.
First, we calculate the normalized difference of the numbers of visible visual elements between query and retrieved visualizations.
Then, similar to \textit{TCR}, we also compute the average \underline{D}ifferences of the numbers of \underline{V}isual \underline{E}lements~($DVE_{ave}$) and standard deviation of differences of the numbers of visual elements~($DVE_{std}$).

Among all metrics, a smaller value of $TCR_{std}$, $DVE_{ave}$ or $DVE_{std}$ is better, since a smaller value of these metrics shows that the method can achieve a more stable performance or smaller difference between the query visualization and the retrieved visualizations. 
On the other hand, a larger value of $TCR_{ave}$ is appreciated, since it demonstrates that the model can retrieve more visualizations of the same type.

\textbf{Results. } Table~\ref{table:quantitative_results} shows the overall results of the four methods.
The results are the average values from five runs of each method.
Our structure-aware approach consistently outperforms other approaches using $TCR_{ave}$ and $TCR_{std}$ and is better than visual information-based methods in $DVE_{ave}$ and $DVE_{std}$.
The results of $DVE_{ave}$s align with the results by Haehn~\textit{et al.}~\cite{haehn2019evaluating}.
They conducted point-cloud experiments and demonstrated that CNNs do not perform well in estimating the difference of numbers of visual elements.
In our experiments, \cnn~also performed worse when using $DVE_{ave}$ as the metric.
Compared with \cnn, both \svg~and \ours~can distinguish the number of visual elements well, which shows the necessity of considering structural information in characterizing the similarity of visualizations. 

\begin{table}[h!]
    \centering
    \small
    \caption{Results of our quantitative evaluation. $TCR$ denotes type-consistent rate and $DVE$ denotes differences in numbers of visual elements. ${ave}$ and $std$ denote average and standard deviation values, respectively. The best results when performing each top-k retrieval are in bold. The results show our structure-aware approach achieves the best performances among the four methods.}
    \Description[]{Results of our quantitative evaluation. $TCR$ denotes type-consistent rate and $DVE$ denotes differences in numbers of visual elements. ${ave}$ and $std$ denote average and standard deviation values, respectively. The best results when performing each top-k retrieval are in bold. The results show our structure-aware approach achieves the best performances among the four methods.}
    \label{table:quantitative_results}
\begin{tabular}{clrrrr}
\toprule
Top-k & Method &  $TCR_{ave}$ &  $TCR_{std}$ &   $DVE_{ave}$ &  $DVE_{std}$  \\
\midrule
\midrule
\multirow{4}{*}{1} 
   & \hog~ &          0.6597 &        0.4738 &     45.1266 & 555.2019  \\
   & \cnn~ &          0.7231 &        0.4474 &      17.1826 & 268.8787  \\
   & \svg~ &          0.7372 &        0.4401 &                \textbf{0.1555} &   \textbf{0.4888}  \\
   & \ours~ &          \textbf{0.7601} &        \textbf{0.4270} &                0.3674 &   1.6752 \\
\cline{1-6}
\multirow{4}{*}{5} 
   & \hog~ &          0.6074 &        0.3249 &    42.9230 & 253.5407  \\
   & \cnn~ &          0.6992 &        0.3150 &               24.4112 & 183.4934  \\
   & \svg~ &          0.7058 &        0.3175 &               \textbf{0.2215} &   \textbf{0.8603} \\
   & \ours~ &          \textbf{0.7383} &        \textbf{0.3096} &     0.6145 &   3.3837  \\
\cline{1-6}
\multirow{4}{*}{10} 
   & \hog~ &          0.5877 &        0.2948  &    46.3089 & 195.6248 \\
   & \cnn~ &          0.6884 &        0.2921  &    26.3127 & 149.2130  \\
   & \svg~ &          0.6893 &        0.2983  &     \textbf{0.2628} &   \textbf{1.0391}  \\
   & \ours~ &          \textbf{0.7246} &        \textbf{0.2900} &                0.7909 &   4.2785 \\
\cline{1-6}
\multirow{4}{*}{20} 
   & \hog~ &          0.5614 &        0.2674  &    49.7751 & 163.7496  \\
   & \cnn~ &          0.6764 &        0.2785  &    25.1575 & 108.1764  \\
   & \svg~ &          0.6713 &        0.2855  &     \textbf{0.3073} &   \textbf{0.9769}  \\
   & \ours~ &          \textbf{0.7076 }&        \textbf{0.2809}  &     1.0432 &   5.1894  \\
\bottomrule
\end{tabular}

\end{table}

\begin{figure*}[h!]
     \centering
     \begin{subfigure}[b]{0.24\linewidth}
         \centering
         \includegraphics[width=\linewidth]{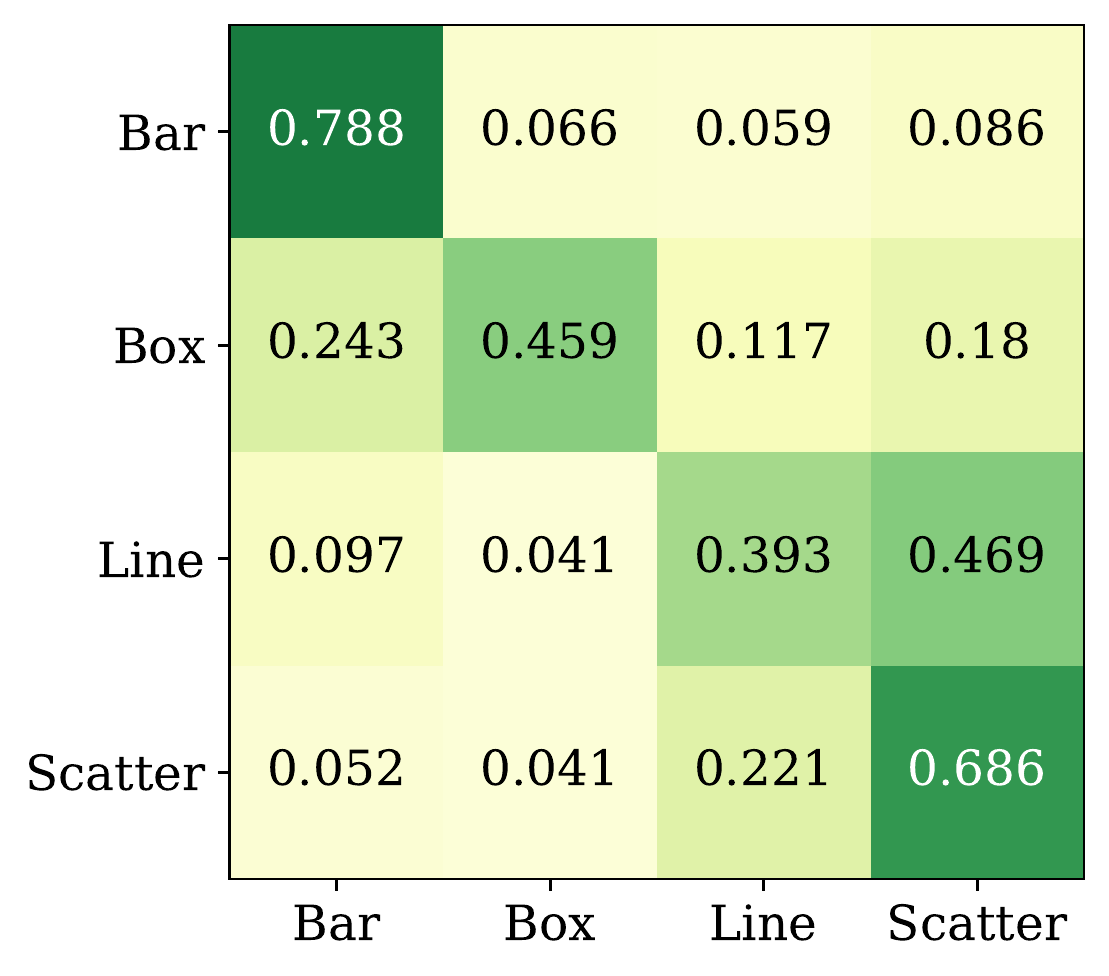}
         \caption{\hog}
     \end{subfigure}
     \hfill
     \begin{subfigure}[b]{0.24\linewidth}
         \centering
         \includegraphics[width=\linewidth]{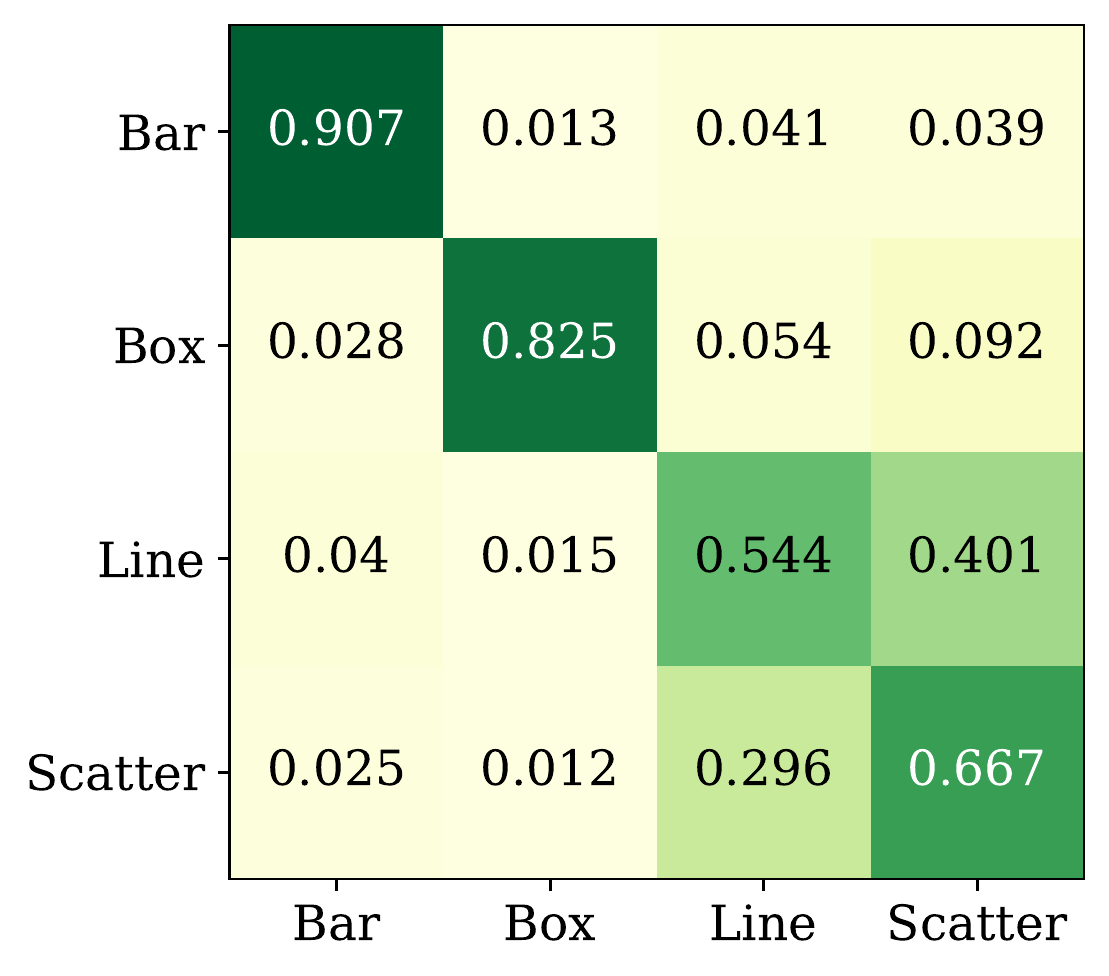}
         \caption{\cnn}
     \end{subfigure}
     \hfill
     \begin{subfigure}[b]{0.24\linewidth}
         \centering
         \includegraphics[width=\linewidth]{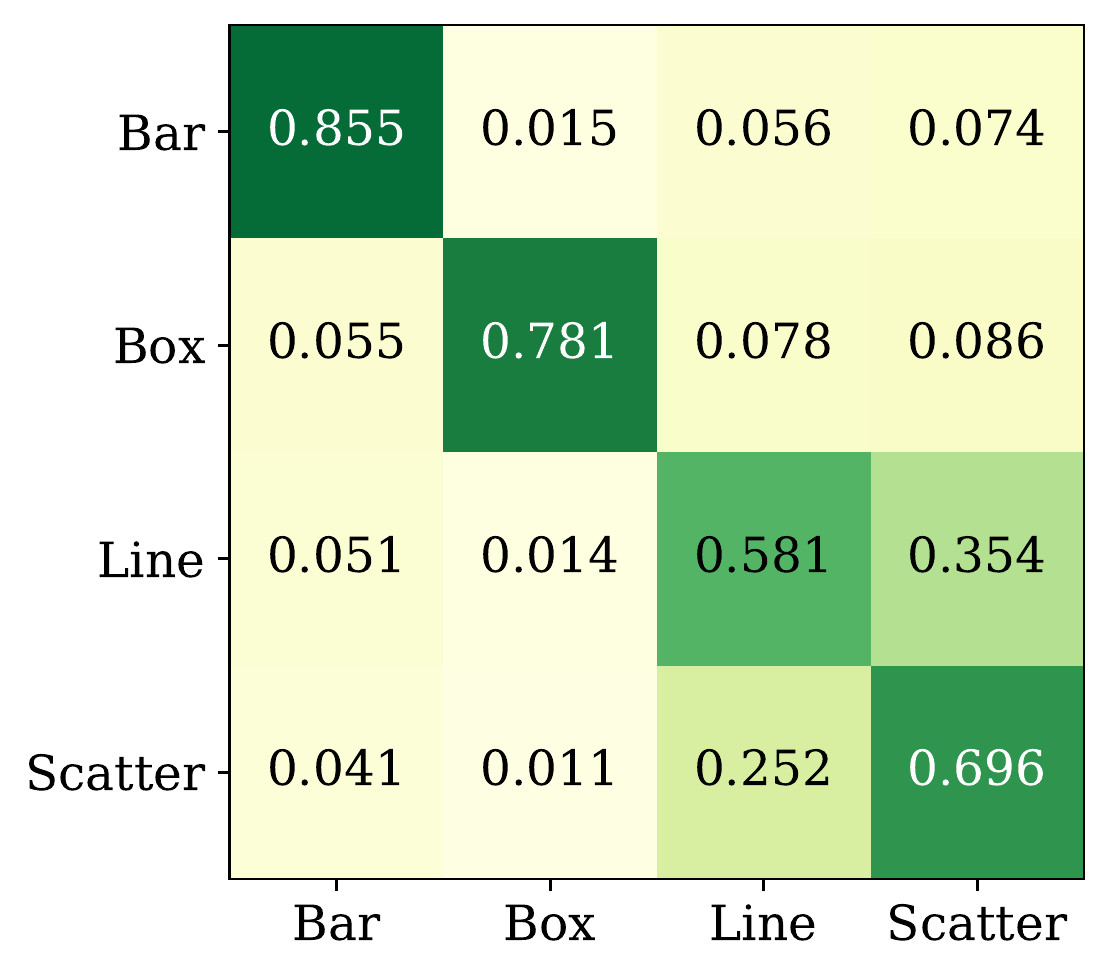}
         \caption{\svg}
     \end{subfigure}
     \hfill
     \begin{subfigure}[b]{0.24\linewidth}
         \centering
         \includegraphics[width=\linewidth]{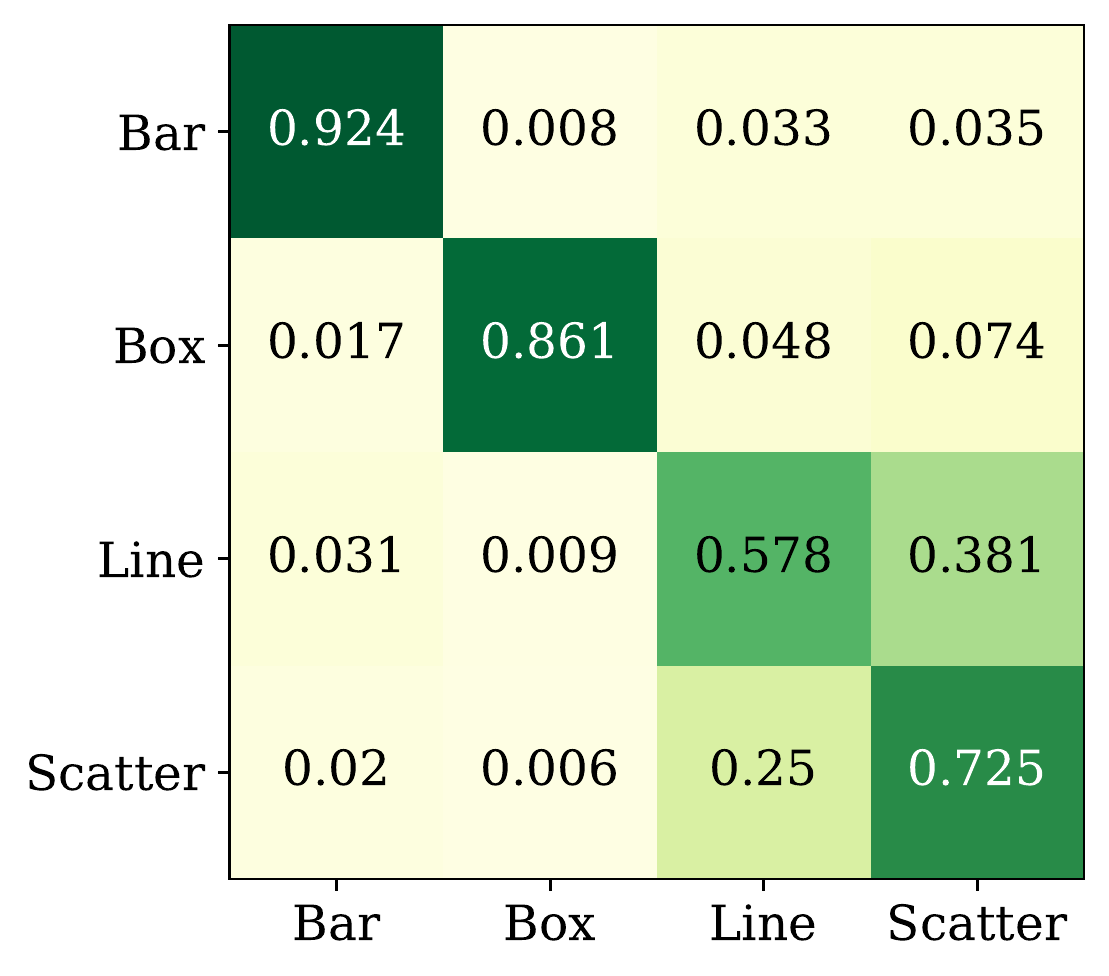}
         \caption{\ours}
     \end{subfigure}
        \caption{\revise{The heatmaps show the $TCR_{ave}$ values of different chart types of four approaches when performing top-5 queries. The horizontal axis denotes the query visualization type and the vertical axis denotes the retrieved visualization type. A block with a darker color indicates a higher $TCR_{ave}$.}}
        \Description[]{The heatmaps show the $TCR_{ave}$ values of different chart types of four approaches when performing top-5 queries. The horizontal axis denotes the query visualization type and the vertical axis denotes the retrieved visualization type. A block with a darker color indicates a higher $TCR_{ave}$. Overall speaking, our approach has the best performance.}
        \label{fig:tcr_each_type}
\end{figure*}

To further understand the pros and cons of different approaches, we investigated the $TCR_{ave}$ values of each visualization type and show the \revise{average confusion matrix in five runs} among all types in Figure~\ref{fig:tcr_each_type}.
From the heatmaps, we can first notice that \hog~is obviously worse than other approaches, probably due to its simplicity by nature.
Thus, in the rest of our paper, we will mainly focus on the comparison among the other three approaches.
\ours~is consistently better than other approaches, which demonstrates the advantages of our structure-aware approach.
An interesting observation is that, though \svg~and \cnn~have close performance in terms of the overall $TCR_{ave}$, they have clear differences in $TCR_{ave}$ of single visualization types.
\cnn~has an advantage on bar charts and box plots while \svg~performs better on line charts and scatter plots.
We checked the detailed results and speculate that the possible reasons are as below.
First, visual elements in bar charts and box plots \revise{often} occupy a large space in the visualization and their shapes are easy to be distinguished by \cnn.
However, lines and scatters are less obvious in visualizations and sometimes can be confused with each other, which may lead to the inaccurate perception by CNNs~(see Figure~\ref{fig:case}(a)).
Compared with \cnn, \svg~explicitly considers all the visual elements as nodes in the graph of visual elements. 
Thus, it performs better on line charts and scatter plots.
The disadvantages of \svg~are mainly shown by the results of box plots and bar charts.
A possible reason is that \svg~only describes \revise{visual elements roughly using their bounding boxes}.
Boxes, bars and scatters can be confused when their bounding boxes are similar.

In summary, the results show that both \cnn~and \svg~have their advantages when dealing with certain types of visualizations.
Thus, the structural information in SVG-based visualizations is essential to be jointly considered with visual information to improve the performance of visualization retrieval.
The results of \ours~in Figure~\ref{fig:tcr_each_type}(d) confirms its effectiveness compared with other approaches.

\subsection{User Study}\label{sec:user_study}
The effectiveness of our approach was further evaluated in a user study with 50 participants recruited from Prolific. 
We asked the participants to score the similarity \revise{between the query visualization and retrieved ones}.
In this section, we introduce the protocol and results of the user study.

\textbf{Dataset.} \revise{In the user study, we first randomly sampled 40 visualizations in the test set.}
For each sampled visualization, we retrieved the top-5 similar visualizations using the four approaches described in Section~\ref{sec:model_setting}.
Then each participant is presented with 5 query visualizations and the corresponding retrieved visualizations for scoring. 
We further added an attention check question to ask the participant to label the similarity between the same visualizations.
\revise{The visualizations used in the user study are available through}  \url{https://structure-vis-retrieval.github.io/}.

\textbf{Procedure.} 
\revise{The user study contains two parts.
Similar to the preliminary study, the first part was to introduce the procedure, collect consent from participants and test the visualization knowledge of the participants.}
After they passed the \revise{test, we moved to the second part. Specifically,} we introduced the criteria for visualization similarity collected in our preliminary study and emphasized that the visualization type, trend of data and number of visual elements are of high priority \revise{based on the results of the preliminary study}.
We also showed several examples to illustrate these criteria for scoring the similarity between visualizations.
\revise{The purpose of introducing these criteria and the corresponding examples is to calibrate the participants' judgment and eliminate the effect of extreme scores given by some participants.}
Then the participants were asked to apply these criteria to score the retrieved \revise{top-5} similar visualizations for each query visualization using a 5-point Likert scale where 1 means the least similar and 5 means the most similar.

\begin{figure}[h!]
    \centering
    \includegraphics[width=0.8\linewidth]{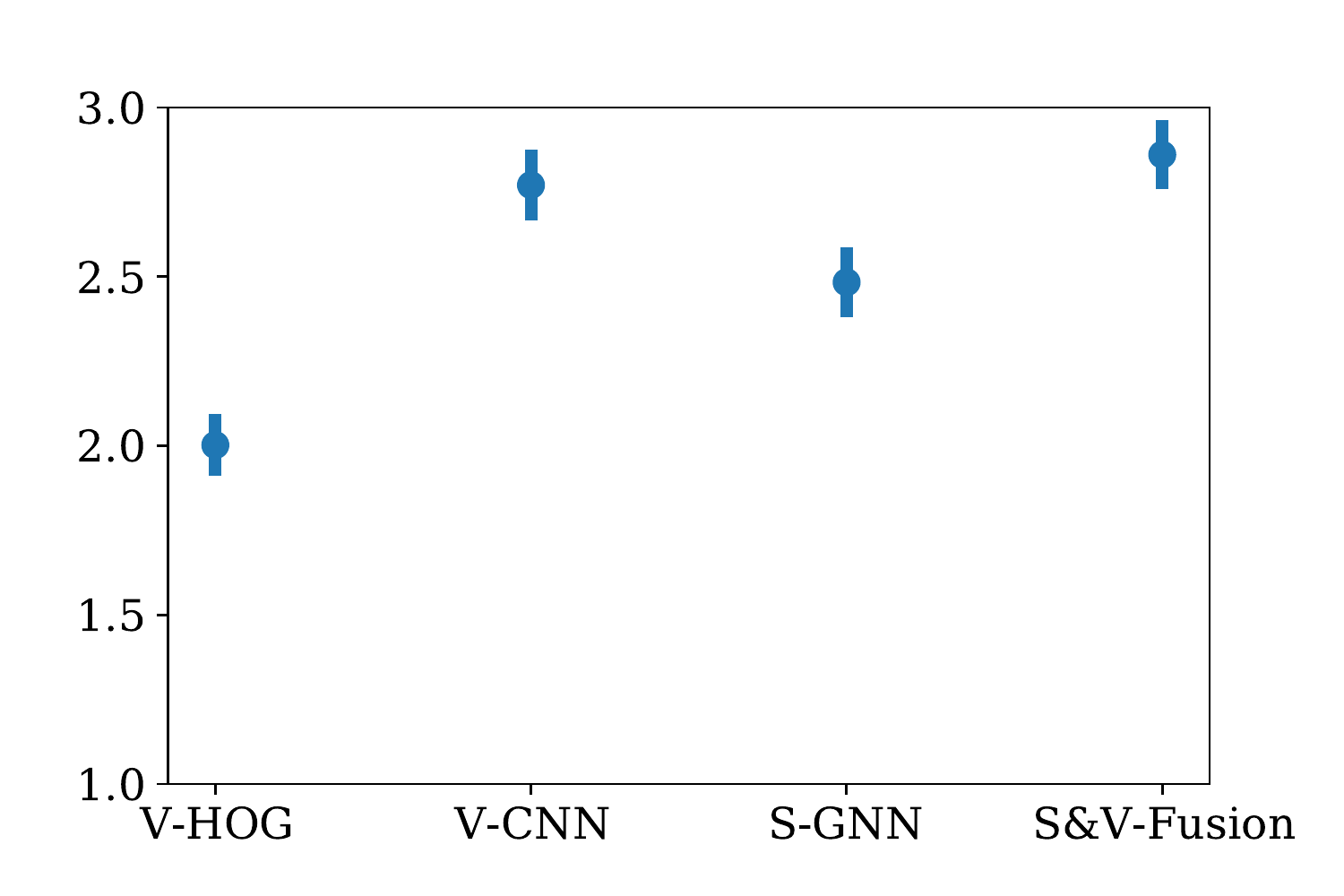}
    \caption{The average similarity scores of retrieved visualizations are shown with 95\% confidence intervals. \ours~outperforms others with statistical significance~($p<0.001$).}
    \Description[]{The average similarity scores of retrieved visualizations are shown with 95\% confidence intervals. Our method outperforms others with statistical significance (p<0.001).}
    \label{fig:user_study}
\end{figure}

\textbf{Results.}
The results of our user study are shown in Figure~\ref{fig:user_study}.
We calculated the average similarity scores of retrieved visualizations by each method~(\ours: 2.8607, \cnn: 2.7707, \svg: 2.4833 and \hog: 2.002). 
A higher score indicates that the retrieved visualizations are more similar to the query one.
According to the results, \ours~outperforms all other models with statistical significance~($p<0.001$) tested with Wilcoxon Signed-rank tests.

\vspace{-1em}

\vspace{-1em}

\subsection{Case Study}\label{sec:case_study}
In this section, we present some examples shown in the user study based on the three most important criteria for visualization similarity.
These cases further illustrate the pros and cons of our structure-aware approach compared with others.
Since \hog~consistently performs worse than others, in this section, we mainly compare \ours, \svg, and \cnn.

\begin{figure*}
    \centering
    \includegraphics[width=\linewidth]{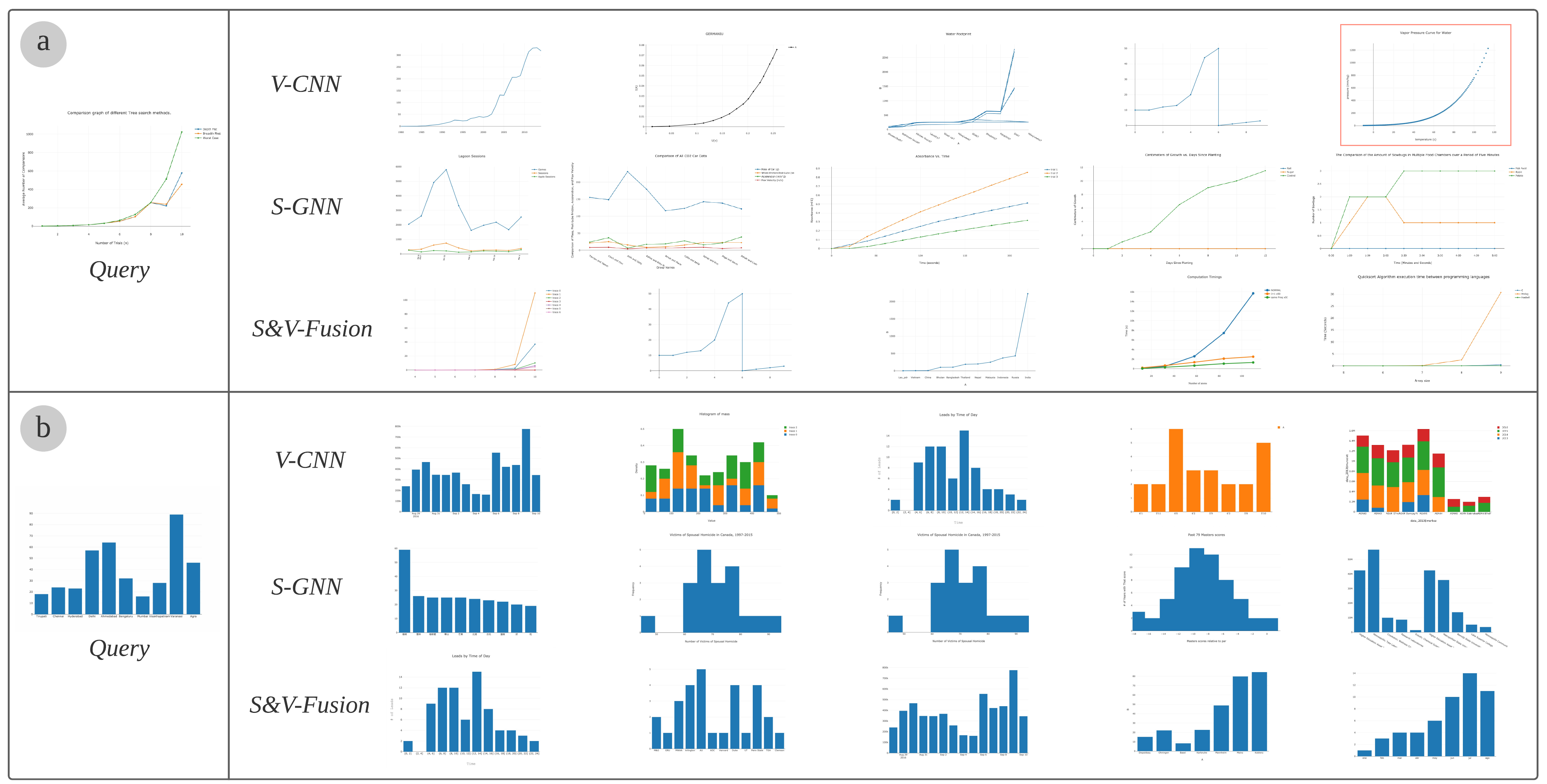}
    \caption{This figure shows examples of query and top-5 retrieved similar visualizations by \cnn, \hog~and \ours.}
    \Description[]{This figure shows examples of query and top-5 retrieved similar visualizations by V-CNN, V-HOG and S&V-Fusion.}
    \label{fig:case}
\end{figure*}

\textbf{Type of Visualizations.} 
As Table~\ref{table:quantitative_results} and Figure~\ref{fig:tcr_each_type} show, \ours~and \svg~have advantages in distinguishing the type of charts, especially between line charts and scatter plots.
A case regarding this phenomenon is identified in our user study, as shown in Figure~\ref{fig:case}(a).
The query visualization in this case is a line chart with an increasing trend.
However, with the representation generated by \cnn, a scatter plot with a similar trend is considered as a similar visualization~(bounded with a red box in Figure~\ref{fig:case}(a)).
Though they share some common features such as the trend of data, they should not be considered as similar visualizations according to the criteria of visualization types\revise{, which is identified in our preliminary study (see Section~\ref{fig:pre_study})}.
Both \ours~and \svg~do not have such kinds of mistakenly retrieved visualizations, which indicates that they can better maintain the visualization type consistency.
Another case in Figure~\ref{fig:case}(b) also reflects a similar issue of \cnn.
\cnn~retrieved multiple stacked bar charts for a plain bar chart while the other two methods only returned similar plain bar charts.

\textbf{Number of Visual Elements.}
The query visualization in Figure~\ref{fig:case}(a) is a line chart with three lines.
\svg~achieves the best results using this criterion since most of the retrieved visualizations have three lines.
Most of the retrieved visualizations by \cnn~are with a single line, which fails to meet the criterion on the numbers of visual elements.

\textbf{Trend of Data.}
Though \svg~shows its advantages in distinguishing chart types and identifying the difference between numbers of visual elements, its capability of considering the trend of data is not as good as \cnn.
As both cases show, almost all the retrieved visualizations by \svg~have trends that are different from the query visualization, which reflects its drawback in effectively representing the trend of data in visualizations.
Thus, 
we considered both the visual and structural information in \ours~to mitigate such an issue.
As both cases show, \ours~achieves a better performance in retrieving visualizations with similar trends while preserving the advantages in the type of visualizations and the number of visual elements.

\section{Discussion}
In this section, we discuss the lessons learned in developing our structure-aware visualization retrieval methods, the generalizability of our approach, 
some potential application scenarios and the limitations of our approach. 
\subsection{Lessons}\label{sec:lessons}
In our approach, we explicitly consider the structural information through deep learning techniques, which enables a new perspective of characterizing the similarity of visualizations.
To facilitate future studies along this direction, we conclude two important lessons, the necessity of structural information in characterizing visualization similarity and deep learning model customization for visualizations.

\textbf{Necessity of Structural Information.}
The similarity between visualizations can be measured through various aspects such as visualization types, colors and trends of data.
Thus, to identify key criteria when determining similarity between visualizations, we conducted a preliminary study with \revise{general} visualization users and summarized three most important criteria, the visualization type, the trend of data and the number of visual elements.
These three criteria indicate the necessity of considering structural information in visualizations since they consider the similarity more at the level of visual elements instead of the level of pixels.
Most of the prior studies merely treat visualizations as \revise{bitmap images} and extract features based on pixels, which ignores the important structural information of visual elements.
Compared with these approaches, our method shows a promising direction of explicitly considering the structural information in SVGs to characterize visualization similarity. 
We treat each visual element in the visualization as a node with basic features in the graph and use edges to represent their spatial and hierarchical relationships, which can effectively reflect the numbers and groups of visual elements.
Since the structural information is mainly at the level of visual elements, we further leverage the pixel-level visual information to make the characterization of visualization more fine-grained.
As our case studies in Section~\ref{sec:case_study} show, \ours~achieves better performance in fulfilling the requirements of retrieving similar visualizations such as the numbers of visual elements and visualization types.

\textbf{Deep Learning Model Customization.}\label{sec:tailored_model}
Deep learning models have been proved to be effective in many applications such as image processing and natural language understanding.
There is also increasing interest in applying deep learning approaches in visualization-related tasks such as automatic visualization generation~\cite{hu2019vizml} and visualization similarity characterization~\cite{zhang2021chartnavigator}.
In our approach, we also apply state-of-the-art contrastive learning models to generate embedding vectors of visualizations.
However, we noticed that these general deep learning models may not be directly applied to visualizations.
For example, as we mentioned in Section~\ref{sec:embedding_bitmap}, flipping is a commonly used operation for data augmentation in general image processing.
However, it is not suitable for visualizations since the trend of data is entirely altered.
Thus, when applying these models, certain customizations should be carefully conducted.
Furthermore, visualization-tailored deep learning models are also desired. 
For example, compared with shape features, the bias of CNN models towards texture features has been observed~\cite{geirhos2018imagenet}.
However, in visualizations, we focus more on the shapes of visual elements instead of the texture.
Thus, a tailored CNN model which emphasizes shape features might be more suitable for visualization-related tasks and can replace the ResNet-50 in our approach.

\subsection{Generalizability and Application Scenarios}
This section discusses the generalizability of our approach and some potential application scenarios of our method. 
The generalizability is from two perspectives: generalizability to visualizations created using other packages~(e.g., D3) and generalizability to multi-view visualizations~(e.g., dashboards or visual analytic systems).

\textbf{Visualizations Created by Other Packages.} 
In our evaluation, we used a visualization corpus crawled from Plotly.
However, our approach is not restricted to visualizations created using Plotly and can be extended to visualizations created using other visualization packages.
\revise{
The requirements of leveraging our approach to retrieve visualizations include a \italic{\textit{unified structure}} of visualizations and the \italic{\textit{consistent usage}} of SVG elements in the visualizations.
Here a unified structure of visualizations requires a consistent way of grouping visual elements.
Our approach does not require a specific criterion of grouping visual elements as long as a unified way of grouping visual elements is applied, for example, grouping \textit{<path>}s by data columns in Plotly.
The consistent usage of SVG elements requires that the same type of SVG elements is employed to
render the same type of visualizations. 
For example, all the bars in bar charts are plotted with \textit{<path>}s.}

\revise{Visualizations created using specification-based packages such as Vega-Lite~\cite{satyanarayan2017vegalite} and Plotly usually fulfill the requirements of our approach.
However, when dealing with the visualizations created with tools with more flexibility, such as Adobe Illustrator\footnote{\url{https://www.adobe.com/products/illustrator.html}} and D3~\cite{bostock2011d3}, 
our approach may meet difficulties due to the inconsistent usage of SVG elements.
For example, when creating visualizations with D3, bars can be created using \textit{<path>}, \textit{<rect>} or \textit{<polygon>} in different visualizations. 
A potential future direction is to improve the generalizability of our structure-aware approach to handle visualizations created with various tools.}

\textbf{Visualizations with Multiple Views.} 
Multi-view visualizations have been widely used to accommodate data with a huge number of attributes~\cite{chen2021multiview}.
Along with its popularity, characterizing multi-view visualizations also attracts researchers' interest~\cite{chen2021multiview, wu2021multivision}.
\revise{Our approach also has the potential to
capture
the structural information of multi-view visualizations by considering more factors, such as the hierarchical and spatial relationship among views.}

\textbf{Application Scenarios.}\label{sec:application}
With the popularity of data visualizations, an emerging research direction is to treat them as a data format and to propose visualization-specific methods for storing, querying and analyzing enormous visualizations~\cite{wu2021survey}.
Our approach has the potential to enable various downstream applications towards this direction.
First, our approach can provide a new way to perform the nearest neighbor query on stored visualizations based on their structural information.
This can also boost the re-use of visualization codes since structural information is highly related to the implementation of a visualization.
Second, a more effective visualization retrieval approach can enhance the large-scale analysis of visualizations.
It allows users to group similar visualizations and conduct further analysis such as understanding \revise{general users'} preferences on visual designs.
Third, it can facilitate the construction of large-scale visualization corpora.
Most of the existing visualization corpora only provide high-level labels of visualizations such as the visualization type~\cite{hu2019vizml} and color usage~\cite{yuan2021colormap}.
With our approach, fine-grained labels~(e.g., the trend of data) of visualizations can be easily labeled based on similar visualization retrieval.

\subsection{Limitations}
The evaluations above have demonstrated the effectiveness of our approach, but it is not without limitations.

\textbf{Evaluation.}
To extensively evaluate our approach, we have conducted a user study, quantitative comparisons and case studies. 
However, there are also limitations in our evaluation.
Among the three most important criteria in evaluating the perceptual similarity of visualizations, We have designed two quantitative metrics to measure the performance of various visualization retrieval methods from the perspective of visualization type consistency and the difference of numbers of visual elements.
However, 
we did not apply a quantitative metric to evaluate the consistency between trends of data in visualizations.
The major reason is that the trends of data in visualizations are complex and difficult to be accurately quantified in a universal way.
For example, in box plots, the trend of data can refer to either the distribution of data in each box or the difference of data between multiple boxes.
\revise{Existing studies also failed to define a universal metric to quantitatively evaluate the trend of data of different types of visualizations (e.g.,~\cite{wilkinson2005graph, wang2017line, kim2021automated}).}
To mitigate the issue that the similarity between data trends in visualizations cannot be quantitatively evaluated,
we further conducted the user study to verify our effectiveness by asking participants to explicitly consider trends of data when scoring the similarity between visualizations.
\revise{Furthermore, as shown in our user study, though our methods outperform other approaches in terms of the 5-point Likert scale, the score (i.e., 2.8607) is still not perfect.
Also, the confusion between line charts and scatter plots can be further reduced.
We suspect the reason is that some line charts are with circles on the data points (see Figure~\ref{fig:case}), which may affect the measurement of similarity.
In the future, more research is expected to improve our structure-aware approach for visualization retrieval, for example, developing visualization-tailored models as mentioned in Section~\ref{sec:tailored_model}.}

\textbf{Feature Fusion.}
In our structure-aware approach for visualization retrieval, we represent the structural and visual information as two embedding vectors.
According to our preliminary comparison with other approaches, we simply concatenate them as the final embedding vector of the visualization~(see Section~\ref{sec:vis_retrieval}).
However, this step is not without limitations.
First, the concatenated embedding vector of a visualization will take additional space to store, which can lead to inefficiency.
Second, the performance of simple concatenation may be limited due to some redundant information inside two embedding vectors, for example, colors.
The redundant information can result in bias when calculating the similarity.
To address the potential drawbacks, it is possible to fuse the visual and structural information based on additional information such as the interaction records used in Screen2Vec~\cite{li2021screen2vec}.
However, due to the lack of such visualization corpus, we have left it as our future work.

\section{Conclusion}
Scalable Vector Graphic (SVG)-based visualizations have been widely used and shared online.
Along with their popularity, retrieving similar SVG-based visualizations has become a critical task.
In this paper, we propose a structure-aware visualization retrieval approach based on the most common perceptual visualization similarity criteria surveyed in our pilot study.
In addition to the widely used visual information in prior studies, our approach further considers the often ignored but essential structural information to advance the performance.
To consider both types of information, we convert SVGs to bitmaps for visual information extraction and graphs for structural information extraction.
Then contrastive representation learning technique 
is employed to generate the low-dimensional embedding vectors of visual and structural information, respectively.
The corresponding visualization is then represented by 
the concatenation of embedding vectors for visualization retrieval.
We conducted extensive evaluations, including quantitative comparisons with prior approaches, a user study and multiple case studies, to demonstrate the effectiveness of our approach.

In future, we would like to extend our structure-aware visualization retrieval approach to more visualization types, e.g., multi-view visualizations.
Another interesting direction is to explore the possibility of directly extracting structural information from bitmap-based visualizations and investigate how our approach can work for the retrieval of bitmap-based visualizations.

\begin{acks}
This research was supported by the Singapore Ministry of Education (MOE) Academic Research Fund (AcRF) Tier 1 grant (Grant number: 20-C220-SMU-011).
Yong Wang is the corresponding author.
We would like to thank Mr. Tianyuan Yao for constructive suggestions, Mr. Jiakai Wang for proofreading and anonymous reviewers for their feedback.
\end{acks}
\balance
\bibliographystyle{ACM-Reference-Format}
\bibliography{reference}


\begin{thebibliography}{59}

%%% the Web address.

\ifx \showCODEN    \undefined \def \showCODEN     #1{\unskip}     \fi
\ifx \showDOI      \undefined \def \showDOI       #1{#1}\fi
\ifx \showISBNx    \undefined \def \showISBNx     #1{\unskip}     \fi
\ifx \showISBNxiii \undefined \def \showISBNxiii  #1{\unskip}     \fi
\ifx \showISSN     \undefined \def \showISSN      #1{\unskip}     \fi
\ifx \showLCCN     \undefined \def \showLCCN      #1{\unskip}     \fi
\ifx \shownote     \undefined \def \shownote      #1{#1}          \fi
\ifx \showarticletitle \undefined \def \showarticletitle #1{#1}   \fi
\ifx \showURL      \undefined \def \showURL       {\relax}        \fi
% invisible to TeX
\providecommand\bibfield[2]{#2}
\providecommand\bibinfo[2]{#2}
\providecommand\natexlab[1]{#1}
\providecommand\showeprint[2][]{arXiv:#2}

\bibitem[\protect\citeauthoryear{Battle, Duan, Miranda, Mukusheva, Chang, and
  Stonebraker}{Battle et~al\mbox{.}}{2018}]%
        {battle2018beagle}
\bibfield{author}{\bibinfo{person}{Leilani Battle}, \bibinfo{person}{Peitong
  Duan}, \bibinfo{person}{Zachery Miranda}, \bibinfo{person}{Dana Mukusheva},
  \bibinfo{person}{Remco Chang}, {and} \bibinfo{person}{Michael Stonebraker}.}
  \bibinfo{year}{2018}\natexlab{}.
\newblock \showarticletitle{Beagle: Automated Extraction and Interpretation of
  Visualizations from the Web}. In \bibinfo{booktitle}{\emph{Proceedings of the
  2018 {CHI} Conference on Human Factors in Computing Systems}}.
  \bibinfo{pages}{1--8}.
\newblock


\bibitem[\protect\citeauthoryear{Bostock, Ogievetsky, and Heer}{Bostock
  et~al\mbox{.}}{2011}]%
        {bostock2011d3}
\bibfield{author}{\bibinfo{person}{Michael Bostock}, \bibinfo{person}{Vadim
  Ogievetsky}, {and} \bibinfo{person}{Jeffrey Heer}.}
  \bibinfo{year}{2011}\natexlab{}.
\newblock \showarticletitle{D{\({^3}\)} Data-Driven Documents}.
\newblock \bibinfo{journal}{\emph{IEEE Transactions on Visualization and
  Computer Graphics}} \bibinfo{volume}{17}, \bibinfo{number}{12}
  (\bibinfo{year}{2011}), \bibinfo{pages}{2301--2309}.
\newblock


\bibitem[\protect\citeauthoryear{Cao, Lin, Guo, Liu, Liu, and Wang}{Cao
  et~al\mbox{.}}{2021}]%
        {cao2021bipartite}
\bibfield{author}{\bibinfo{person}{Jiangxia Cao}, \bibinfo{person}{Xixun Lin},
  \bibinfo{person}{Shu Guo}, \bibinfo{person}{Luchen Liu},
  \bibinfo{person}{Tingwen Liu}, {and} \bibinfo{person}{Bin Wang}.}
  \bibinfo{year}{2021}\natexlab{}.
\newblock \showarticletitle{Bipartite Graph Embedding via Mutual Information
  Maximization}. In \bibinfo{booktitle}{\emph{Proceedings of the 14th ACM
  International Conference on Web Search and Data Mining}}.
  \bibinfo{pages}{635--643}.
\newblock


\bibitem[\protect\citeauthoryear{Chen, Kornblith, Norouzi, and Hinton}{Chen
  et~al\mbox{.}}{2020}]%
        {chen2020simclr}
\bibfield{author}{\bibinfo{person}{Ting Chen}, \bibinfo{person}{Simon
  Kornblith}, \bibinfo{person}{Mohammad Norouzi}, {and}
  \bibinfo{person}{Geoffrey Hinton}.} \bibinfo{year}{2020}\natexlab{}.
\newblock \showarticletitle{A Simple Framework for Contrastive Learning of
  Visual Representations}. In \bibinfo{booktitle}{\emph{Proceedings of the 37th
  International Conference on Machine Learning}}. \bibinfo{pages}{1597--1607}.
\newblock


\bibitem[\protect\citeauthoryear{Chen and He}{Chen and He}{2021}]%
        {chen2020simsiam}
\bibfield{author}{\bibinfo{person}{Xinlei Chen} {and} \bibinfo{person}{Kaiming
  He}.} \bibinfo{year}{2021}\natexlab{}.
\newblock \showarticletitle{Exploring Simple Siamese Representation Learning}.
  In \bibinfo{booktitle}{\emph{Proceedings of the 2021 IEEE/CVF Conference on
  Computer Vision and Pattern Recognition}}. \bibinfo{pages}{15750--15758}.
\newblock


\bibitem[\protect\citeauthoryear{Chen, Zeng, Lin, AI-maneea, Roberts, and
  Chang}{Chen et~al\mbox{.}}{2021}]%
        {chen2021multiview}
\bibfield{author}{\bibinfo{person}{Xi Chen}, \bibinfo{person}{Wei Zeng},
  \bibinfo{person}{Yanna Lin}, \bibinfo{person}{Hayder~Mahdi AI-maneea},
  \bibinfo{person}{Jonathan Roberts}, {and} \bibinfo{person}{Remco Chang}.}
  \bibinfo{year}{2021}\natexlab{}.
\newblock \showarticletitle{Composition and Configuration Patterns in
  Multiple-View Visualizations}.
\newblock \bibinfo{journal}{\emph{IEEE Transactions on Visualization and
  Computer Graphics}} \bibinfo{volume}{27}, \bibinfo{number}{2}
  (\bibinfo{year}{2021}), \bibinfo{pages}{1514--1524}.
\newblock


\bibitem[\protect\citeauthoryear{Chen, Cafarella, and Adar}{Chen
  et~al\mbox{.}}{2015}]%
        {chen2015diagramflyer}
\bibfield{author}{\bibinfo{person}{Zhe Chen}, \bibinfo{person}{Michael
  Cafarella}, {and} \bibinfo{person}{Eytan Adar}.}
  \bibinfo{year}{2015}\natexlab{}.
\newblock \showarticletitle{DiagramFlyer: A Search Engine for Data-Driven
  Diagrams}. In \bibinfo{booktitle}{\emph{Proceedings of the 24th International
  Conference on World Wide Web}}. \bibinfo{pages}{183--186}.
\newblock


\bibitem[\protect\citeauthoryear{Cleveland}{Cleveland}{1979}]%
        {cleveland1979loess}
\bibfield{author}{\bibinfo{person}{William~S. Cleveland}.}
  \bibinfo{year}{1979}\natexlab{}.
\newblock \showarticletitle{Robust Locally Weighted Regression and Smoothing
  Scatterplots}.
\newblock \bibinfo{journal}{\emph{Journal of the American Statistical
  Association}} \bibinfo{volume}{74}, \bibinfo{number}{368}
  (\bibinfo{year}{1979}), \bibinfo{pages}{829--836}.
\newblock


\bibitem[\protect\citeauthoryear{Dalal and Triggs}{Dalal and Triggs}{2005}]%
        {dalal2005hog}
\bibfield{author}{\bibinfo{person}{Navneet Dalal} {and} \bibinfo{person}{Bill
  Triggs}.} \bibinfo{year}{2005}\natexlab{}.
\newblock \showarticletitle{Histograms of Oriented Gradients for Human
  Detection}. In \bibinfo{booktitle}{\emph{Proceedings of the 2005 IEEE
  Computer Society Conference on Computer Vision and Pattern Recognition}},
  Vol.~\bibinfo{volume}{1}. \bibinfo{pages}{886--893}.
\newblock


\bibitem[\protect\citeauthoryear{Deng, Dong, Socher, Li, Li, and Fei-Fei}{Deng
  et~al\mbox{.}}{2009}]%
        {deng2009imagenet}
\bibfield{author}{\bibinfo{person}{Jia Deng}, \bibinfo{person}{Wei Dong},
  \bibinfo{person}{Richard Socher}, \bibinfo{person}{Li-Jia Li},
  \bibinfo{person}{Kai Li}, {and} \bibinfo{person}{Li Fei-Fei}.}
  \bibinfo{year}{2009}\natexlab{}.
\newblock \showarticletitle{ImageNet: A Large-scale Hierarchical Image
  Database}. In \bibinfo{booktitle}{\emph{Proceedings of the 2009 IEEE/CVF
  Conference on Computer Vision and Pattern Recognition}}.
  \bibinfo{pages}{248--255}.
\newblock


\bibitem[\protect\citeauthoryear{Feng, Wang, and Li}{Feng
  et~al\mbox{.}}{2014}]%
        {feng2014cross}
\bibfield{author}{\bibinfo{person}{Fangxiang Feng}, \bibinfo{person}{Xiaojie
  Wang}, {and} \bibinfo{person}{Ruifan Li}.} \bibinfo{year}{2014}\natexlab{}.
\newblock \showarticletitle{Cross-modal Retrieval with Correspondence
  Autoencoder}. In \bibinfo{booktitle}{\emph{Proceedings of the 22nd ACM
  International Conference on Multimedia}}. \bibinfo{pages}{7--16}.
\newblock


\bibitem[\protect\citeauthoryear{Fu, Zhu, Cui, Ge, Wang, Zhang, Huang, Tang,
  Zhang, and Ma}{Fu et~al\mbox{.}}{2021}]%
        {chartem2021fu}
\bibfield{author}{\bibinfo{person}{Jiayun Fu}, \bibinfo{person}{Bin Zhu},
  \bibinfo{person}{Weiwei Cui}, \bibinfo{person}{Song Ge}, \bibinfo{person}{Yun
  Wang}, \bibinfo{person}{Haidong Zhang}, \bibinfo{person}{He Huang},
  \bibinfo{person}{Yuanyuan Tang}, \bibinfo{person}{Dongmei Zhang}, {and}
  \bibinfo{person}{Xiaojing Ma}.} \bibinfo{year}{2021}\natexlab{}.
\newblock \showarticletitle{Chartem: Reviving Chart Images with Data
  Embedding}.
\newblock \bibinfo{journal}{\emph{IEEE Transactions on Visualization and
  Computer Graphics}} \bibinfo{volume}{27}, \bibinfo{number}{2}
  (\bibinfo{year}{2021}), \bibinfo{pages}{337--346}.
\newblock


\bibitem[\protect\citeauthoryear{Geirhos, Rubisch, Michaelis, Bethge, Wichmann,
  and Brendel}{Geirhos et~al\mbox{.}}{2018}]%
        {geirhos2018imagenet}
\bibfield{author}{\bibinfo{person}{Robert Geirhos}, \bibinfo{person}{Patricia
  Rubisch}, \bibinfo{person}{Claudio Michaelis}, \bibinfo{person}{Matthias
  Bethge}, \bibinfo{person}{Felix~A Wichmann}, {and} \bibinfo{person}{Wieland
  Brendel}.} \bibinfo{year}{2018}\natexlab{}.
\newblock \showarticletitle{ImageNet-trained CNNs are Biased towards Texture;
  Increasing Shape Bias Improves Accuracy and Robustness}.
\newblock \bibinfo{journal}{\emph{CoRR}}  \bibinfo{volume}{abs/1811.12231}
  (\bibinfo{year}{2018}).
\newblock


\bibitem[\protect\citeauthoryear{Grill, Strub, Altch{\'e}, Tallec, Richemond,
  Buchatskaya, Doersch, Pires, Guo, Azar, et~al\mbox{.}}{Grill
  et~al\mbox{.}}{2020}]%
        {grill2020byol}
\bibfield{author}{\bibinfo{person}{Jean-Bastien Grill},
  \bibinfo{person}{Florian Strub}, \bibinfo{person}{Florent Altch{\'e}},
  \bibinfo{person}{Corentin Tallec}, \bibinfo{person}{Pierre~H Richemond},
  \bibinfo{person}{Elena Buchatskaya}, \bibinfo{person}{Carl Doersch},
  \bibinfo{person}{Bernardo~Avila Pires}, \bibinfo{person}{Zhaohan~Daniel Guo},
  \bibinfo{person}{Mohammad~Gheshlaghi Azar}, {et~al\mbox{.}}}
  \bibinfo{year}{2020}\natexlab{}.
\newblock \showarticletitle{Bootstrap Your Own Latent: A New Approach to
  Self-Supervised Learning}.
\newblock \bibinfo{journal}{\emph{CoRR}}  \bibinfo{volume}{abs/2006.07733}
  (\bibinfo{year}{2020}).
\newblock


\bibitem[\protect\citeauthoryear{Haehn, Tompkin, and Pfister}{Haehn
  et~al\mbox{.}}{2019}]%
        {haehn2019evaluating}
\bibfield{author}{\bibinfo{person}{Daniel Haehn}, \bibinfo{person}{James
  Tompkin}, {and} \bibinfo{person}{Hanspeter Pfister}.}
  \bibinfo{year}{2019}\natexlab{}.
\newblock \showarticletitle{Evaluating `Graphical Perception' with CNNs}.
\newblock \bibinfo{journal}{\emph{IEEE Transactions on Visualization and
  Computer Graphics}} \bibinfo{volume}{25}, \bibinfo{number}{1}
  (\bibinfo{year}{2019}), \bibinfo{pages}{641--650}.
\newblock


\bibitem[\protect\citeauthoryear{He, Zhang, Ren, and Sun}{He
  et~al\mbox{.}}{2015}]%
        {he2015resnet}
\bibfield{author}{\bibinfo{person}{Kaiming He}, \bibinfo{person}{Xiangyu
  Zhang}, \bibinfo{person}{Shaoqing Ren}, {and} \bibinfo{person}{Jian Sun}.}
  \bibinfo{year}{2015}\natexlab{}.
\newblock \showarticletitle{Deep Residual Learning for Image Recognition}.
\newblock \bibinfo{journal}{\emph{CoRR}}  \bibinfo{volume}{abs/1512.03385}
  (\bibinfo{year}{2015}).
\newblock


\bibitem[\protect\citeauthoryear{Hoque and Agrawala}{Hoque and
  Agrawala}{2020}]%
        {hoque2020searching}
\bibfield{author}{\bibinfo{person}{Enamul Hoque} {and} \bibinfo{person}{Maneesh
  Agrawala}.} \bibinfo{year}{2020}\natexlab{}.
\newblock \showarticletitle{Searching the Visual Style and Structure of {D3}
  Visualizations}.
\newblock \bibinfo{journal}{\emph{IEEE Transactions on Visualization and
  Computer Graphics}} \bibinfo{volume}{26}, \bibinfo{number}{1}
  (\bibinfo{year}{2020}), \bibinfo{pages}{1236--1245}.
\newblock


\bibitem[\protect\citeauthoryear{Hu, Bakker, Li, Kraska, and Hidalgo}{Hu
  et~al\mbox{.}}{2019}]%
        {hu2019vizml}
\bibfield{author}{\bibinfo{person}{Kevin~Zeng Hu}, \bibinfo{person}{Michiel~A.
  Bakker}, \bibinfo{person}{Stephen Li}, \bibinfo{person}{Tim Kraska}, {and}
  \bibinfo{person}{C{\'{e}}sar~A. Hidalgo}.} \bibinfo{year}{2019}\natexlab{}.
\newblock \showarticletitle{VizML: {A} Machine Learning Approach to
  Visualization Recommendation}. In \bibinfo{booktitle}{\emph{Proceedings of
  the 2019 {CHI} Conference on Human Factors in Computing Systems}}.
  \bibinfo{pages}{1--12}.
\newblock


\bibitem[\protect\citeauthoryear{Jaiswal, Babu, Zadeh, Banerjee, and
  Makedon}{Jaiswal et~al\mbox{.}}{2020}]%
        {jaiswal2020contrastive}
\bibfield{author}{\bibinfo{person}{Ashish Jaiswal},
  \bibinfo{person}{Ashwin~Ramesh Babu}, \bibinfo{person}{Mohammad~Zaki Zadeh},
  \bibinfo{person}{Debapriya Banerjee}, {and} \bibinfo{person}{Fillia
  Makedon}.} \bibinfo{year}{2020}\natexlab{}.
\newblock \showarticletitle{A Survey on Contrastive Self-supervised Learning}.
\newblock \bibinfo{journal}{\emph{CoRR}}  \bibinfo{volume}{abs/2011.00362}
  (\bibinfo{year}{2020}).
\newblock


\bibitem[\protect\citeauthoryear{Jung, Kim, Song, Hwang, Lee, Kim, and
  Seo}{Jung et~al\mbox{.}}{2017}]%
        {jung2017chartsense}
\bibfield{author}{\bibinfo{person}{Daekyoung Jung}, \bibinfo{person}{Wonjae
  Kim}, \bibinfo{person}{Hyunjoo Song}, \bibinfo{person}{Jeong-in Hwang},
  \bibinfo{person}{Bongshin Lee}, \bibinfo{person}{Bohyoung Kim}, {and}
  \bibinfo{person}{Jinwook Seo}.} \bibinfo{year}{2017}\natexlab{}.
\newblock \showarticletitle{ChartSense: Interactive Data Extraction from Chart
  Images}. In \bibinfo{booktitle}{\emph{Proceedings of the 2017 {CHI}
  Conference on Human Factors in Computing Systems}}.
  \bibinfo{pages}{6706--6717}.
\newblock


\bibitem[\protect\citeauthoryear{Kim, Rossi, Sarma, Moritz, and Hullman}{Kim
  et~al\mbox{.}}{2021}]%
        {kim2021automated}
\bibfield{author}{\bibinfo{person}{Hyeok Kim}, \bibinfo{person}{Ryan Rossi},
  \bibinfo{person}{Abhraneel Sarma}, \bibinfo{person}{Dominik Moritz}, {and}
  \bibinfo{person}{Jessica Hullman}.} \bibinfo{year}{2021}\natexlab{}.
\newblock \showarticletitle{An Automated Approach to Reasoning About
  Task-Oriented Insights in Responsive Visualization}.
\newblock \bibinfo{journal}{\emph{CoRR}}  \bibinfo{volume}{abs/2107.08141}
  (\bibinfo{year}{2021}).
\newblock


\bibitem[\protect\citeauthoryear{Li, Wang, Zhang, Song, and Qu}{Li
  et~al\mbox{.}}{2021b}]%
        {li2021kg4vis}
\bibfield{author}{\bibinfo{person}{Haotian Li}, \bibinfo{person}{Yong Wang},
  \bibinfo{person}{Songheng Zhang}, \bibinfo{person}{Yangqiu Song}, {and}
  \bibinfo{person}{Huamin Qu}.} \bibinfo{year}{2021}\natexlab{b}.
\newblock \showarticletitle{KG4Vis: {A} Knowledge Graph-Based Approach for
  Visualization Recommendation}.
\newblock \bibinfo{journal}{\emph{CoRR}}  \bibinfo{volume}{abs/2107.12548}
  (\bibinfo{year}{2021}).
\newblock


\bibitem[\protect\citeauthoryear{Li, Wei, Wang, Song, and Qu}{Li
  et~al\mbox{.}}{2020}]%
        {li2020peer}
\bibfield{author}{\bibinfo{person}{Haotian Li}, \bibinfo{person}{Huan Wei},
  \bibinfo{person}{Yong Wang}, \bibinfo{person}{Yangqiu Song}, {and}
  \bibinfo{person}{Huamin Qu}.} \bibinfo{year}{2020}\natexlab{}.
\newblock \showarticletitle{Peer-inspired Student Performance Prediction in
  Interactive Online Question Pools with Graph Neural Network}. In
  \bibinfo{booktitle}{\emph{Proceedings of the 29th ACM International
  Conference on Information \& Knowledge Management}}.
  \bibinfo{pages}{2589--2596}.
\newblock


\bibitem[\protect\citeauthoryear{Li, Popowski, Mitchell, and Myers}{Li
  et~al\mbox{.}}{2021a}]%
        {li2021screen2vec}
\bibfield{author}{\bibinfo{person}{Toby~Jia{-}Jun Li}, \bibinfo{person}{Lindsay
  Popowski}, \bibinfo{person}{Tom~M. Mitchell}, {and} \bibinfo{person}{Brad~A.
  Myers}.} \bibinfo{year}{2021}\natexlab{a}.
\newblock \showarticletitle{Screen2Vec: Semantic Embedding of {GUI} Screens and
  {GUI} Components}. In \bibinfo{booktitle}{\emph{Proceedings of the 2021 {CHI}
  Conference on Human Factors in Computing Systems}}. \bibinfo{pages}{1--15}.
\newblock


\bibitem[\protect\citeauthoryear{Li, Carberry, Fang, McCoy, Peterson, and
  Stagitis}{Li et~al\mbox{.}}{2015}]%
        {li2015novel}
\bibfield{author}{\bibinfo{person}{Zhuo Li}, \bibinfo{person}{Sandra Carberry},
  \bibinfo{person}{Hui Fang}, \bibinfo{person}{Kathleen~F McCoy},
  \bibinfo{person}{Kelly Peterson}, {and} \bibinfo{person}{Matthew Stagitis}.}
  \bibinfo{year}{2015}\natexlab{}.
\newblock \showarticletitle{A Novel Methodology for Retrieving Infographics
  Utilizing Structure and Message Content}.
\newblock \bibinfo{journal}{\emph{Data \& Knowledge Engineering}}
  \bibinfo{volume}{100} (\bibinfo{year}{2015}), \bibinfo{pages}{191--210}.
\newblock


\bibitem[\protect\citeauthoryear{Ma, Tung, Wang, Gao, Pan, and Chen}{Ma
  et~al\mbox{.}}{2020}]%
        {ma2020scatternet}
\bibfield{author}{\bibinfo{person}{Yuxin Ma}, \bibinfo{person}{Anthony K.~H.
  Tung}, \bibinfo{person}{Wei Wang}, \bibinfo{person}{Xiang Gao},
  \bibinfo{person}{Zhigeng Pan}, {and} \bibinfo{person}{Wei Chen}.}
  \bibinfo{year}{2020}\natexlab{}.
\newblock \showarticletitle{ScatterNet: {A} Deep Subjective Similarity Model
  for Visual Analysis of Scatterplots}.
\newblock \bibinfo{journal}{\emph{IEEE Transactions on Visualization and
  Computer Graphics}} \bibinfo{volume}{26}, \bibinfo{number}{3}
  (\bibinfo{year}{2020}), \bibinfo{pages}{1562--1576}.
\newblock


\bibitem[\protect\citeauthoryear{Ngiam, Khosla, Kim, Nam, Lee, and Ng}{Ngiam
  et~al\mbox{.}}{2011}]%
        {ngiam2011multimodal}
\bibfield{author}{\bibinfo{person}{Jiquan Ngiam}, \bibinfo{person}{Aditya
  Khosla}, \bibinfo{person}{Mingyu Kim}, \bibinfo{person}{Juhan Nam},
  \bibinfo{person}{Honglak Lee}, {and} \bibinfo{person}{Andrew~Y. Ng}.}
  \bibinfo{year}{2011}\natexlab{}.
\newblock \showarticletitle{Multimodal Deep Learning}. In
  \bibinfo{booktitle}{\emph{Proceedings of the 28th International Conference on
  International Conference on Machine Learning}}. \bibinfo{pages}{689--696}.
\newblock


\bibitem[\protect\citeauthoryear{Oppermann, Kincaid, and Munzner}{Oppermann
  et~al\mbox{.}}{2021}]%
        {oppermann2021vizcommender}
\bibfield{author}{\bibinfo{person}{Michael Oppermann}, \bibinfo{person}{Robert
  Kincaid}, {and} \bibinfo{person}{Tamara Munzner}.}
  \bibinfo{year}{2021}\natexlab{}.
\newblock \showarticletitle{VizCommender: Computing Text-Based Similarity in
  Visualization Repositories for Content-Based Recommendations}.
\newblock \bibinfo{journal}{\emph{IEEE Transactions on Visualization and
  Computer Graphics}} \bibinfo{volume}{27}, \bibinfo{number}{2}
  (\bibinfo{year}{2021}), \bibinfo{pages}{495--505}.
\newblock


\bibitem[\protect\citeauthoryear{Oppermann and Munzner}{Oppermann and
  Munzner}{2021}]%
        {opperman2021vizsnippets}
\bibfield{author}{\bibinfo{person}{Michael Oppermann} {and}
  \bibinfo{person}{Tamara Munzner}.} \bibinfo{year}{2021}\natexlab{}.
\newblock \showarticletitle{VizSnippets: Compressing Visualization Bundles Into
  Representative Previews for Browsing Visualization Collections}.
\newblock \bibinfo{journal}{\emph{IEEE Transactions on Visualization and
  Computer Graphics}} (\bibinfo{year}{2021}).
\newblock


\bibitem[\protect\citeauthoryear{Patil, Li, Fisher, Savva, and Zhang}{Patil
  et~al\mbox{.}}{2021}]%
        {patil2021layoutgmn}
\bibfield{author}{\bibinfo{person}{Akshay~Gadi Patil}, \bibinfo{person}{Manyi
  Li}, \bibinfo{person}{Matthew Fisher}, \bibinfo{person}{Manolis Savva}, {and}
  \bibinfo{person}{Hao Zhang}.} \bibinfo{year}{2021}\natexlab{}.
\newblock \showarticletitle{LayoutGMN: Neural Graph Matching for Structural
  Layout Similarity}. In \bibinfo{booktitle}{\emph{Proceedings of the IEEE/CVF
  Conference on Computer Vision and Pattern Recognition}}.
  \bibinfo{pages}{11048--11057}.
\newblock


\bibitem[\protect\citeauthoryear{Qian, Sun, Cui, Lou, Zhang, and Zhang}{Qian
  et~al\mbox{.}}{2020}]%
        {qian2020retrieve}
\bibfield{author}{\bibinfo{person}{Chunyao Qian}, \bibinfo{person}{Shizhao
  Sun}, \bibinfo{person}{Weiwei Cui}, \bibinfo{person}{Jian-Guang Lou},
  \bibinfo{person}{Haidong Zhang}, {and} \bibinfo{person}{Dongmei Zhang}.}
  \bibinfo{year}{2020}\natexlab{}.
\newblock \showarticletitle{Retrieve-Then-Adapt: Example-based Automatic
  Generation for Proportion-related Infographics}.
\newblock \bibinfo{journal}{\emph{IEEE Transactions on Visualization and
  Computer Graphics}} \bibinfo{volume}{27}, \bibinfo{number}{2}
  (\bibinfo{year}{2020}), \bibinfo{pages}{443--452}.
\newblock


\bibitem[\protect\citeauthoryear{Raji, Duncan, Hobson, and Huang}{Raji
  et~al\mbox{.}}{2021}]%
        {raji2021dataless}
\bibfield{author}{\bibinfo{person}{Mohammad Raji}, \bibinfo{person}{Jeremiah
  Duncan}, \bibinfo{person}{Tanner Hobson}, {and} \bibinfo{person}{Jian
  Huang}.} \bibinfo{year}{2021}\natexlab{}.
\newblock \showarticletitle{Dataless Sharing of Interactive Visualization}.
\newblock \bibinfo{journal}{\emph{IEEE Transactions on Visualization and
  Computer Graphics}} \bibinfo{volume}{27}, \bibinfo{number}{9}
  (\bibinfo{year}{2021}), \bibinfo{pages}{3656--3669}.
\newblock


\bibitem[\protect\citeauthoryear{Saleh, Dontcheva, Hertzmann, and Liu}{Saleh
  et~al\mbox{.}}{2015}]%
        {saleh2015similarity}
\bibfield{author}{\bibinfo{person}{Babak Saleh}, \bibinfo{person}{Mira
  Dontcheva}, \bibinfo{person}{Aaron Hertzmann}, {and}
  \bibinfo{person}{Zhicheng Liu}.} \bibinfo{year}{2015}\natexlab{}.
\newblock \showarticletitle{Learning Style Similarity for Searching
  Infographics}. In \bibinfo{booktitle}{\emph{Proceedings of the 41st Graphics
  Interface Conference}}. \bibinfo{pages}{59--64}.
\newblock


\bibitem[\protect\citeauthoryear{Satyanarayan, Lee, Ren, Heer, Stasko,
  Thompson, Brehmer, and Liu}{Satyanarayan et~al\mbox{.}}{2020}]%
        {satyanarayan2020critical}
\bibfield{author}{\bibinfo{person}{Arvind Satyanarayan},
  \bibinfo{person}{Bongshin Lee}, \bibinfo{person}{Donghao Ren},
  \bibinfo{person}{Jeffrey Heer}, \bibinfo{person}{John~T. Stasko},
  \bibinfo{person}{John Thompson}, \bibinfo{person}{Matthew Brehmer}, {and}
  \bibinfo{person}{Zhicheng Liu}.} \bibinfo{year}{2020}\natexlab{}.
\newblock \showarticletitle{Critical Reflections on Visualization Authoring
  Systems}.
\newblock \bibinfo{journal}{\emph{IEEE Transactions on Visualization and
  Computer Graphics}} \bibinfo{volume}{26}, \bibinfo{number}{1}
  (\bibinfo{year}{2020}), \bibinfo{pages}{461--471}.
\newblock


\bibitem[\protect\citeauthoryear{Satyanarayan, Moritz, Wongsuphasawat, and
  Heer}{Satyanarayan et~al\mbox{.}}{2017}]%
        {satyanarayan2017vegalite}
\bibfield{author}{\bibinfo{person}{Arvind Satyanarayan},
  \bibinfo{person}{Dominik Moritz}, \bibinfo{person}{Kanit Wongsuphasawat},
  {and} \bibinfo{person}{Jeffrey Heer}.} \bibinfo{year}{2017}\natexlab{}.
\newblock \showarticletitle{Vega-Lite: A Grammar of Interactive Graphics}.
\newblock \bibinfo{journal}{\emph{IEEE Transactions on Visualization and
  Computer Graphics}} \bibinfo{volume}{23}, \bibinfo{number}{1}
  (\bibinfo{year}{2017}), \bibinfo{pages}{341--350}.
\newblock


\bibitem[\protect\citeauthoryear{Savva, Kong, Chhajta, Fei-Fei, Agrawala, and
  Heer}{Savva et~al\mbox{.}}{2011}]%
        {savva2011revision}
\bibfield{author}{\bibinfo{person}{Manolis Savva}, \bibinfo{person}{Nicholas
  Kong}, \bibinfo{person}{Arti Chhajta}, \bibinfo{person}{Li Fei-Fei},
  \bibinfo{person}{Maneesh Agrawala}, {and} \bibinfo{person}{Jeffrey Heer}.}
  \bibinfo{year}{2011}\natexlab{}.
\newblock \showarticletitle{ReVision: Automated Classification, Analysis and
  Redesign of Chart Images}. In \bibinfo{booktitle}{\emph{Proceedings of the
  24th ACM Symposium on User Interface Software and Technology}}.
  \bibinfo{pages}{393--402}.
\newblock


\bibitem[\protect\citeauthoryear{Schlichtkrull, Kipf, Bloem, Van Den~Berg,
  Titov, and Welling}{Schlichtkrull et~al\mbox{.}}{2018}]%
        {schlichtkrull2018modeling}
\bibfield{author}{\bibinfo{person}{Michael Schlichtkrull},
  \bibinfo{person}{Thomas~N Kipf}, \bibinfo{person}{Peter Bloem},
  \bibinfo{person}{Rianne Van Den~Berg}, \bibinfo{person}{Ivan Titov}, {and}
  \bibinfo{person}{Max Welling}.} \bibinfo{year}{2018}\natexlab{}.
\newblock \showarticletitle{Modeling Relational Data with Graph Convolutional
  Networks}. In \bibinfo{booktitle}{\emph{Proceedings of the 2018 European
  Semantic Web Conference}}. \bibinfo{pages}{593--607}.
\newblock


\bibitem[\protect\citeauthoryear{Siddiqui, Kim, Lee, Karahalios, and
  Parameswaran}{Siddiqui et~al\mbox{.}}{2016}]%
        {siddiqui2016effortless}
\bibfield{author}{\bibinfo{person}{Tarique Siddiqui}, \bibinfo{person}{Albert
  Kim}, \bibinfo{person}{John Lee}, \bibinfo{person}{Karrie Karahalios}, {and}
  \bibinfo{person}{Aditya Parameswaran}.} \bibinfo{year}{2016}\natexlab{}.
\newblock \showarticletitle{Effortless Data Exploration with zenvisage: An
  Expressive and Interactive Visual Analytics System}.
\newblock \bibinfo{journal}{\emph{Proceedings of the VLDB Endowment}}
  \bibinfo{volume}{10}, \bibinfo{number}{4} (\bibinfo{year}{2016}).
\newblock


\bibitem[\protect\citeauthoryear{Siddiqui, Luh, Wang, Karahalios, and
  Parameswaran}{Siddiqui et~al\mbox{.}}{2020}]%
        {siddiqui2020shapesearch}
\bibfield{author}{\bibinfo{person}{Tarique Siddiqui}, \bibinfo{person}{Paul
  Luh}, \bibinfo{person}{Zesheng Wang}, \bibinfo{person}{Karrie Karahalios},
  {and} \bibinfo{person}{Aditya Parameswaran}.}
  \bibinfo{year}{2020}\natexlab{}.
\newblock \showarticletitle{ShapeSearch: A Flexible and Efficient System for
  Shape-based Exploration of Trendlines}. In
  \bibinfo{booktitle}{\emph{Proceedings of the 2020 ACM SIGMOD International
  Conference on Management of Data}}. \bibinfo{pages}{51--65}.
\newblock


\bibitem[\protect\citeauthoryear{Siegel, Horvitz, Levin, Divvala, and
  Farhadi}{Siegel et~al\mbox{.}}{2016}]%
        {siegel2016figureseer}
\bibfield{author}{\bibinfo{person}{Noah Siegel}, \bibinfo{person}{Zachary
  Horvitz}, \bibinfo{person}{Roie Levin}, \bibinfo{person}{Santosh Divvala},
  {and} \bibinfo{person}{Ali Farhadi}.} \bibinfo{year}{2016}\natexlab{}.
\newblock \showarticletitle{FigureSeer: Parsing Result-Figures in Research
  Papers}. In \bibinfo{booktitle}{\emph{Proceedings of the 14th European
  Conference on Computer Vision}}. \bibinfo{pages}{664--680}.
\newblock


\bibitem[\protect\citeauthoryear{Stitz, Gratzl, Piringer, Zichner, and
  Streit}{Stitz et~al\mbox{.}}{2018}]%
        {stitz2018knowledgepearls}
\bibfield{author}{\bibinfo{person}{Holger Stitz}, \bibinfo{person}{Samuel
  Gratzl}, \bibinfo{person}{Harald Piringer}, \bibinfo{person}{Thomas Zichner},
  {and} \bibinfo{person}{Marc Streit}.} \bibinfo{year}{2018}\natexlab{}.
\newblock \showarticletitle{KnowledgePearls: Provenance-Based Visualization
  Retrieval}.
\newblock \bibinfo{journal}{\emph{IEEE Transactions on Visualization and
  Computer Graphics}} \bibinfo{volume}{25}, \bibinfo{number}{1}
  (\bibinfo{year}{2018}), \bibinfo{pages}{120--130}.
\newblock


\bibitem[\protect\citeauthoryear{Sun, Hoffman, Verma, and Tang}{Sun
  et~al\mbox{.}}{2019}]%
        {sun2019infograph}
\bibfield{author}{\bibinfo{person}{Fan-Yun Sun}, \bibinfo{person}{Jordan
  Hoffman}, \bibinfo{person}{Vikas Verma}, {and} \bibinfo{person}{Jian Tang}.}
  \bibinfo{year}{2019}\natexlab{}.
\newblock \showarticletitle{InfoGraph: Unsupervised and Semi-supervised
  Graph-Level Representation Learning via Mutual Information Maximization}. In
  \bibinfo{booktitle}{\emph{Proceedings of the 10th International Conference on
  Learning Representations}}.
\newblock


\bibitem[\protect\citeauthoryear{Tkachev, Frey, and Ertl}{Tkachev
  et~al\mbox{.}}{2021}]%
        {tkachev2021s4}
\bibfield{author}{\bibinfo{person}{Gleb Tkachev}, \bibinfo{person}{Steffen
  Frey}, {and} \bibinfo{person}{Thomas Ertl}.} \bibinfo{year}{2021}\natexlab{}.
\newblock \showarticletitle{S4: Self-Supervised learning of Spatiotemporal
  Similarity}.
\newblock \bibinfo{journal}{\emph{IEEE Transactions on Visualization and
  Computer Graphics}} (\bibinfo{year}{2021}).
\newblock


\bibitem[\protect\citeauthoryear{Vartak, Rahman, Madden, Parameswaran, and
  Polyzotis}{Vartak et~al\mbox{.}}{2015}]%
        {vartak2015seedb}
\bibfield{author}{\bibinfo{person}{Manasi Vartak}, \bibinfo{person}{Sajjadur
  Rahman}, \bibinfo{person}{Samuel Madden}, \bibinfo{person}{Aditya
  Parameswaran}, {and} \bibinfo{person}{Neoklis Polyzotis}.}
  \bibinfo{year}{2015}\natexlab{}.
\newblock \showarticletitle{SeeDB: Efficient Data-Driven Visualization
  Recommendations to Support Visual Analytics}.
\newblock \bibinfo{journal}{\emph{Proceedings of the VLDB Endowment}}
  \bibinfo{volume}{8}, \bibinfo{number}{13} (\bibinfo{year}{2015}),
  \bibinfo{pages}{2182--2193}.
\newblock


\bibitem[\protect\citeauthoryear{Velickovic, Fedus, Hamilton, Li{\`{o}},
  Bengio, and Hjelm}{Velickovic et~al\mbox{.}}{2018}]%
        {velivckovic2018deep}
\bibfield{author}{\bibinfo{person}{Petar Velickovic}, \bibinfo{person}{William
  Fedus}, \bibinfo{person}{William~L. Hamilton}, \bibinfo{person}{Pietro
  Li{\`{o}}}, \bibinfo{person}{Yoshua Bengio}, {and} \bibinfo{person}{R.~Devon
  Hjelm}.} \bibinfo{year}{2018}\natexlab{}.
\newblock \showarticletitle{Deep Graph Infomax}.
\newblock \bibinfo{journal}{\emph{CoRR}}  \bibinfo{volume}{abs/1809.10341}
  (\bibinfo{year}{2018}).
\newblock


\bibitem[\protect\citeauthoryear{Wang, Chen, Wang, and Qu}{Wang
  et~al\mbox{.}}{2021}]%
        {wang2021survey}
\bibfield{author}{\bibinfo{person}{Qianwen Wang}, \bibinfo{person}{Zhutian
  Chen}, \bibinfo{person}{Yong Wang}, {and} \bibinfo{person}{Huamin Qu}.}
  \bibinfo{year}{2021}\natexlab{}.
\newblock \showarticletitle{A Survey on ML4VIS: Applying Machine Learning
  Advances to Data Visualization}.
\newblock \bibinfo{journal}{\emph{IEEE Transactions on Visualization and
  Computer Graphics}} (\bibinfo{year}{2021}).
\newblock


\bibitem[\protect\citeauthoryear{Wang, Han, Zhu, Deussen, and Chen}{Wang
  et~al\mbox{.}}{2017}]%
        {wang2017line}
\bibfield{author}{\bibinfo{person}{Yunhai Wang}, \bibinfo{person}{Fubo Han},
  \bibinfo{person}{Lifeng Zhu}, \bibinfo{person}{Oliver Deussen}, {and}
  \bibinfo{person}{Baoquan Chen}.} \bibinfo{year}{2017}\natexlab{}.
\newblock \showarticletitle{Line Graph or Scatter Plot? Automatic Selection of
  Methods for Visualizing Trends in Time Series}.
\newblock \bibinfo{journal}{\emph{IEEE Transactions on Visualization and
  Computer graphics}} \bibinfo{volume}{24}, \bibinfo{number}{2}
  (\bibinfo{year}{2017}), \bibinfo{pages}{1141--1154}.
\newblock


\bibitem[\protect\citeauthoryear{Wang, Jin, Wang, Cui, Ma, and Qu}{Wang
  et~al\mbox{.}}{2019}]%
        {wang2019deepdrawing}
\bibfield{author}{\bibinfo{person}{Yong Wang}, \bibinfo{person}{Zhihua Jin},
  \bibinfo{person}{Qianwen Wang}, \bibinfo{person}{Weiwei Cui},
  \bibinfo{person}{Tengfei Ma}, {and} \bibinfo{person}{Huamin Qu}.}
  \bibinfo{year}{2019}\natexlab{}.
\newblock \showarticletitle{DeepDrawing: A Deep Learning Approach to Graph
  Drawing}.
\newblock \bibinfo{journal}{\emph{IEEE Transactions on Visualization and
  Computer Graphics}} \bibinfo{volume}{26}, \bibinfo{number}{1}
  (\bibinfo{year}{2019}), \bibinfo{pages}{676--686}.
\newblock


\bibitem[\protect\citeauthoryear{Wilkinson, Anand, and Grossman}{Wilkinson
  et~al\mbox{.}}{2005}]%
        {wilkinson2005graph}
\bibfield{author}{\bibinfo{person}{Leland Wilkinson}, \bibinfo{person}{Anushka
  Anand}, {and} \bibinfo{person}{Robert Grossman}.}
  \bibinfo{year}{2005}\natexlab{}.
\newblock \showarticletitle{Graph-theoretic Scagnostics}. In
  \bibinfo{booktitle}{\emph{Proceedings of the 2005 IEEE Symposium on
  Information Visualization}}. \bibinfo{pages}{21--21}.
\newblock


\bibitem[\protect\citeauthoryear{Wu, Tong, Dwyer, Lee, Isenberg, and Qu}{Wu
  et~al\mbox{.}}{2020}]%
        {wu2020mobilevisfixer}
\bibfield{author}{\bibinfo{person}{Aoyu Wu}, \bibinfo{person}{Wai Tong},
  \bibinfo{person}{Tim Dwyer}, \bibinfo{person}{Bongshin Lee},
  \bibinfo{person}{Petra Isenberg}, {and} \bibinfo{person}{Huamin Qu}.}
  \bibinfo{year}{2020}\natexlab{}.
\newblock \showarticletitle{MobileVisFixer: Tailoring Web Visualizations for
  Mobile Phones Leveraging an Explainable Reinforcement Learning Framework}.
\newblock \bibinfo{journal}{\emph{IEEE Transactions on Visualization and
  Computer Graphics}} \bibinfo{volume}{27}, \bibinfo{number}{2}
  (\bibinfo{year}{2020}), \bibinfo{pages}{464--474}.
\newblock


\bibitem[\protect\citeauthoryear{Wu, Wang, Shu, Moritz, Cui, Zhang, Zhang, and
  Qu}{Wu et~al\mbox{.}}{2021b}]%
        {wu2021survey}
\bibfield{author}{\bibinfo{person}{Aoyu Wu}, \bibinfo{person}{Yun Wang},
  \bibinfo{person}{Xinhuan Shu}, \bibinfo{person}{Dominik Moritz},
  \bibinfo{person}{Weiwei Cui}, \bibinfo{person}{Haidong Zhang},
  \bibinfo{person}{Dongmei Zhang}, {and} \bibinfo{person}{Huamin Qu}.}
  \bibinfo{year}{2021}\natexlab{b}.
\newblock \showarticletitle{Survey on Artificial Intelligence Approaches for
  Visualization Data}.
\newblock \bibinfo{journal}{\emph{CoRR}}  \bibinfo{volume}{abs/2102.01330}
  (\bibinfo{year}{2021}).
\newblock


\bibitem[\protect\citeauthoryear{Wu, Wang, Zhou, He, Zhang, Qu, and Zhang}{Wu
  et~al\mbox{.}}{2021c}]%
        {wu2021multivision}
\bibfield{author}{\bibinfo{person}{Aoyu Wu}, \bibinfo{person}{Yun Wang},
  \bibinfo{person}{Mengyu Zhou}, \bibinfo{person}{Xinyi He},
  \bibinfo{person}{Haidong Zhang}, \bibinfo{person}{Huamin Qu}, {and}
  \bibinfo{person}{Dongmei Zhang}.} \bibinfo{year}{2021}\natexlab{c}.
\newblock \showarticletitle{MultiVision: Designing Analytical Dashboards with
  Deep Learning Based Recommendation}.
\newblock \bibinfo{journal}{\emph{CoRR}}  \bibinfo{volume}{abs/2107.07823}
  (\bibinfo{year}{2021}).
\newblock


\bibitem[\protect\citeauthoryear{Wu, Lin, Tan, Gao, and Li}{Wu
  et~al\mbox{.}}{2021a}]%
        {wu2021self}
\bibfield{author}{\bibinfo{person}{Lirong Wu}, \bibinfo{person}{Haitao Lin},
  \bibinfo{person}{Cheng Tan}, \bibinfo{person}{Zhangyang Gao}, {and}
  \bibinfo{person}{Stan~Z Li}.} \bibinfo{year}{2021}\natexlab{a}.
\newblock \showarticletitle{Self-supervised Learning on Graphs: Contrastive,
  Generative, or Predictive}.
\newblock \bibinfo{journal}{\emph{IEEE Transactions on Knowledge and Data
  Engineering}} (\bibinfo{year}{2021}).
\newblock


\bibitem[\protect\citeauthoryear{Xu, Hu, Leskovec, and Jegelka}{Xu
  et~al\mbox{.}}{2018}]%
        {xu2018gin}
\bibfield{author}{\bibinfo{person}{Keyulu Xu}, \bibinfo{person}{Weihua Hu},
  \bibinfo{person}{Jure Leskovec}, {and} \bibinfo{person}{Stefanie Jegelka}.}
  \bibinfo{year}{2018}\natexlab{}.
\newblock \showarticletitle{How Powerful are Graph Neural Networks?}
\newblock \bibinfo{journal}{\emph{CoRR}}  \bibinfo{volume}{abs/1810.00826}
  (\bibinfo{year}{2018}).
\newblock


\bibitem[\protect\citeauthoryear{Yuan, Zeng, Fu, Zeng, Li, Fu, and Qu}{Yuan
  et~al\mbox{.}}{2021}]%
        {yuan2021colormap}
\bibfield{author}{\bibinfo{person}{Linping Yuan}, \bibinfo{person}{Wei Zeng},
  \bibinfo{person}{Siwei Fu}, \bibinfo{person}{Zhiliang Zeng},
  \bibinfo{person}{Haotian Li}, \bibinfo{person}{Chi{-}Wing Fu}, {and}
  \bibinfo{person}{Huamin Qu}.} \bibinfo{year}{2021}\natexlab{}.
\newblock \showarticletitle{Deep Colormap Extraction from Visualizations}.
\newblock \bibinfo{journal}{\emph{CoRR}}  \bibinfo{volume}{abs/2103.00741}
  (\bibinfo{year}{2021}).
\newblock


\bibitem[\protect\citeauthoryear{Zhang, Li, and Wang}{Zhang
  et~al\mbox{.}}{2021b}]%
        {zhang2021viscode}
\bibfield{author}{\bibinfo{person}{Peiying Zhang}, \bibinfo{person}{Chenhui
  Li}, {and} \bibinfo{person}{Changbo Wang}.} \bibinfo{year}{2021}\natexlab{b}.
\newblock \showarticletitle{VisCode: Embedding Information in Visualization
  Images using Encoder-Decoder Network}.
\newblock \bibinfo{journal}{\emph{IEEE Transactions on Visualization and
  Computer Graphics}} \bibinfo{volume}{27}, \bibinfo{number}{2}
  (\bibinfo{year}{2021}), \bibinfo{pages}{326--336}.
\newblock


\bibitem[\protect\citeauthoryear{Zhang, Feng, Chen, Chen, Zheng, Luo, Huang,
  and Tung}{Zhang et~al\mbox{.}}{2021a}]%
        {zhang2021chartnavigator}
\bibfield{author}{\bibinfo{person}{Tianye Zhang}, \bibinfo{person}{Haozhe
  Feng}, \bibinfo{person}{Wei Chen}, \bibinfo{person}{Zexian Chen},
  \bibinfo{person}{Wenting Zheng}, \bibinfo{person}{Xiao-Nan Luo},
  \bibinfo{person}{Wenqi Huang}, {and} \bibinfo{person}{Anthony~KH Tung}.}
  \bibinfo{year}{2021}\natexlab{a}.
\newblock \showarticletitle{ChartNavigator: An Interactive Pattern
  Identification and Annotation Framework for Charts}.
\newblock \bibinfo{journal}{\emph{IEEE Transactions on Knowledge and Data
  Engineering}} (\bibinfo{year}{2021}).
\newblock


\bibitem[\protect\citeauthoryear{Zhao, Fan, and Feng}{Zhao
  et~al\mbox{.}}{2020}]%
        {zhao2020chartseer}
\bibfield{author}{\bibinfo{person}{Jian Zhao}, \bibinfo{person}{Mingming Fan},
  {and} \bibinfo{person}{Mi Feng}.} \bibinfo{year}{2020}\natexlab{}.
\newblock \showarticletitle{ChartSeer: Interactive Steering Exploratory Visual
  Analysis with Machine Intelligence}.
\newblock \bibinfo{journal}{\emph{IEEE Transactions on Visualization and
  Computer Graphics}} (\bibinfo{year}{2020}).
\newblock


\bibitem[\protect\citeauthoryear{Zhou, Cui, Hu, Zhang, Yang, Liu, Wang, Li, and
  Sun}{Zhou et~al\mbox{.}}{2020}]%
        {zhou2020graph}
\bibfield{author}{\bibinfo{person}{Jie Zhou}, \bibinfo{person}{Ganqu Cui},
  \bibinfo{person}{Shengding Hu}, \bibinfo{person}{Zhengyan Zhang},
  \bibinfo{person}{Cheng Yang}, \bibinfo{person}{Zhiyuan Liu},
  \bibinfo{person}{Lifeng Wang}, \bibinfo{person}{Changcheng Li}, {and}
  \bibinfo{person}{Maosong Sun}.} \bibinfo{year}{2020}\natexlab{}.
\newblock \showarticletitle{Graph Neural Networks: A Review of Methods and
  Applications}.
\newblock \bibinfo{journal}{\emph{AI Open}}  \bibinfo{volume}{1}
  (\bibinfo{year}{2020}), \bibinfo{pages}{57--81}.
\newblock


\end{thebibliography}

\end{document}